\documentclass[aps,twoside,prb,twocolumn,showpacs,showkeys,citeautoscript]{revtex4-1}
\usepackage{graphicx}
\usepackage{amsmath,amssymb,amstext}
\usepackage{datetime}

\newcommand{\dd}{d}
\newcommand{\bvec}[1]{\boldsymbol{#1}}
\newcommand{\ket}[1]{\left| #1 \,\right\rangle}
\newcommand{\bra}[1]{\left\langle #1 \,\right|}

\newcommand{\nket}[1]{| #1 \,\rangle}
\newcommand{\nbra}[1]{\langle #1 \,|}

\newcommand{\lsco}{LSCO}

\begin{document}

\title{Numerical Study of Charge Transport of Overdoped La$_{2-x}$Sr$_{x}$CuO$_{4}$ within Semiclassical Boltzmann Transport Theory}

\author{Jonathan M. \surname{Buhmann}}
\affiliation{Institute for Theoretical Physics, ETH Z\"{u}rich, 8093 Z\"{u}rich, Switzerland}
\author{Matthias \surname{Ossadnik}}
\affiliation{Institute for Theoretical Physics, ETH Z\"{u}rich, 8093 Z\"{u}rich, Switzerland}
\author{T. M. \surname{Rice}}
\affiliation{Institute for Theoretical Physics, ETH Z\"{u}rich, 8093 Z\"{u}rich, Switzerland}
\author{Manfred \surname{Sigrist}}
\affiliation{Institute for Theoretical Physics, ETH Z\"{u}rich, 8093 Z\"{u}rich, Switzerland}

\begin{abstract}
     The in-plane resistivity of the high-temperature oxide superconductor La$_{2-x}$Sr$_{x}$CuO$_{4}$ [\lsco]  shows a strong growth of a contribution linear in temperature as the doping is reduced in the overdoped region toward optimal. This linear term is a signature of non-Fermi liquid behavior. We find  that the appearance of a linear term in the resistivity can arise in a semiclassical Boltzmann transport theory which uses renormalized quasiparticle scattering rates derived in a functional renormalization group calculation and an empirical band structure fitted to angle-resolved photoemission spectroscopy data on \lsco. The linearized Boltzmann equation is solved numerically by discretizing the Brillouin zone in a way that fits best to the Fermi surface geometry. The main trends in the development of the anomalous temperature dependence are well reproduced. There is a substantial underestimation of the magnitude of the resistivity which is expected in view of the moderate to weak values we chose for the onsite repulsion to stay within the one-loop renormalization group approximation. The analysis was extended to the Seebeck coefficient with similar agreement with the main trends in the data.
     \begin{center}
     This article is accepted for publication (Jan. 2nd, 2013): Phys. Rev. B \textbf{87}, 035129 (2013)\\
     \url{http://link.aps.org/doi/10.1103/PhysRevB.87.035129}
     \end{center}
\end{abstract}


\maketitle

\section{Introduction}\label{sec: introduction}

The unconventional temperature scaling of the in plane resistivity in high-$T_{c}$ cuprates has been of considerable interest for many years. It is now widely accepted that overdoped cuprates exhibit a $T$-linear resistivity at high temperature which crosses over to a $T^{2}$ dependence at lower temperature [\onlinecite{hussey2008,tyler1997,yoshida1999,manako1992,mackenzie1996,naqiba2003}]. The onset of the linear temperature dependence appears to coincide with the onset of superconductivity and  the crossover temperature to a conventional  $T^{2}$ dependence drops monotonically with the doping level to zero, roughly around optimal doping [\onlinecite{hussey2009,hussey2011,greven2012,nakamae2003}].  Hussey and collaborators measured the resistivity on a number of cuprates particularly the single layer Tl2201 in the overdoped region [\onlinecite{french2009,abdel2007}] and LSCO into the underdoped region [\onlinecite{hussey2009,hussey2011}]. Very recently the Bari\v{s}i\'{c} et al. [\onlinecite{greven2012}] reported results for the resistivity for more cuprates, e.g. Hg1201 and YBCO over a wider density range. They found that when the resistivity was normalized to a give a resistance per Cu$_{4}$O$_{4}$ plaquette, $\rho_{\square}$, universal results for all cuprates followed with identical coefficients.  Further they argued for a universal trend in both the linear $(\rho_{\square}\approx A_{1\square}T)$ and quadratic $(\rho_{\square}\approx A_{2\square}T^{2})$ regimes, with the coefficients $A_{1}$ and $A_{2}$ proportional to $1/p$, the inverse hole density. As Bari\v{s}i\'{c} et al. [\onlinecite{greven2012}] remarked, if one uses a Drude formula for resistance, this behavior is indicative of a carrier density that is proportional to the density of holes rather than electrons. Such behavior is consistent with a doped Mott insulator scenario. Anderson [\onlinecite{anderson1,anderson2,anderson3}] has argued that the cuprates as doped Mott insulators should be in the strong coupling regime of the Hubbard model at all hole densities. He has put forward the Hidden Fermi Liquid [HFL] ansatz to describe the connection between the Fermi surface before Gutzwiller projection and the electronic state that results from the removal of doubly occupied states by Gutzwiller projection. This HFL ansatz gives a resistivity with the observed temperature dependence, crossing over from a linear to a quadratic form as the temperature is lowered. 

An earlier viewpoint interprets the cuprate phase diagram in terms of an underlying quantum critical point separating the overdoped and underdoped regions. In these theories the effective interactions between electrons are mediated and controlled by the order parameter fluctuations associated with the quantum critical point [\onlinecite{QCPinCupratePhaseDiagram1,QCPinCupratePhaseDiagram2,senthil2004,abanov2003}]. Another approach recently put forward by Kokalj and coworkers [\onlinecite{kokalj2011,kokalj2012}]  is based on an anisotropic marginal Fermi liquid self energy. Their phenomenological theory gives good fits to the resistivity and also to the angular dependent magnetoresistance [ADMR] results [\onlinecite{abdel2006}] and a number of other properties.

\begin{figure*}[t]

\includegraphics[width=0.95\textwidth]{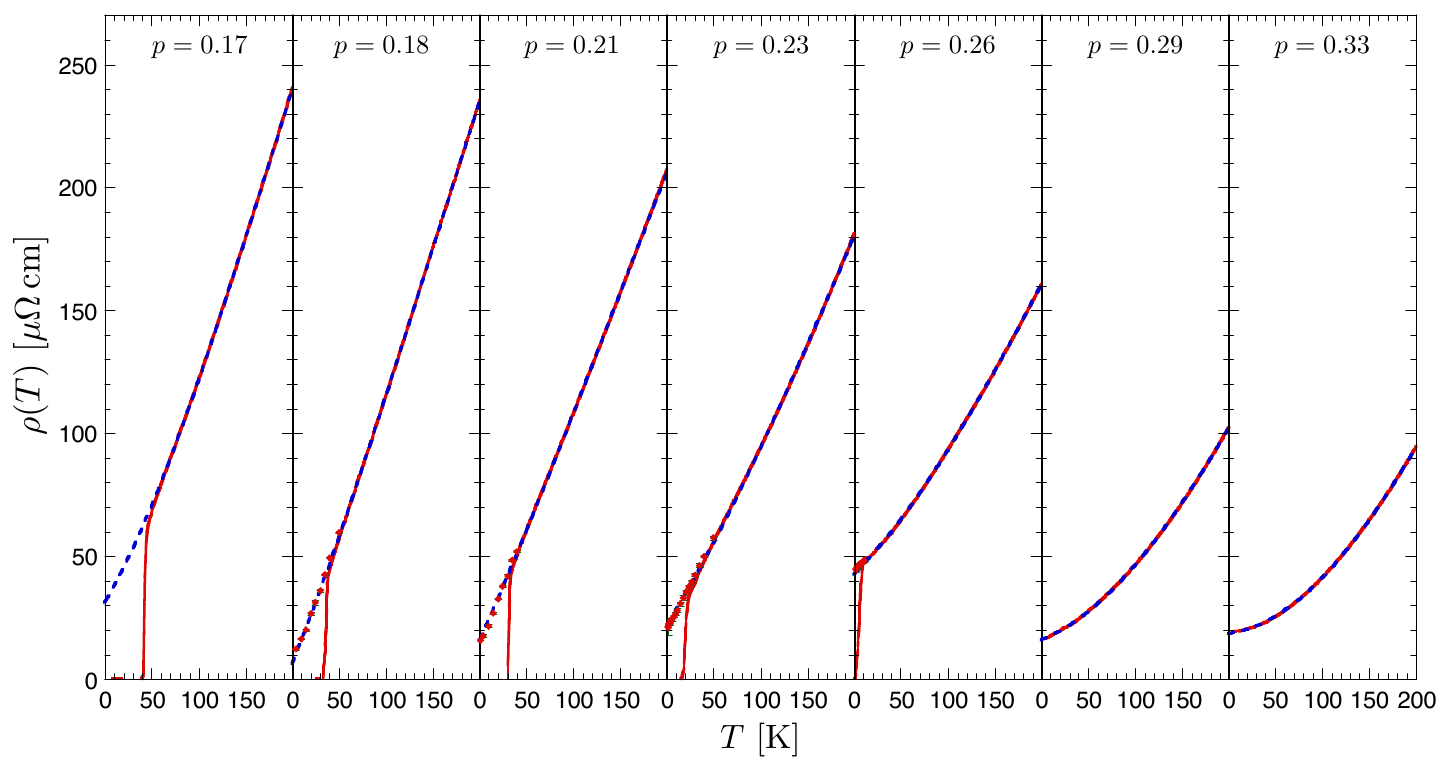}
\caption{Experimental results of Cooper et al. [\onlinecite{hussey2009}]. The in-plane resistivity of \lsco\ is plotted as a function of the temperature for samples with doping levels from $p=0.17$ to $p=0.33$. The red line represents data without external magnetic field, cf. the superconducting transition, while the red diamonds represent data in high magnetic field which suppresses the superconductivity. The blue dotted line is a fit with to the normal state resistivity extrapolated down to zero temperature.}
\label{fig: experimental resistivity}

\end{figure*}

In an earlier study, which employed a functional renormalization group [FRG] treatment of the two-dimensional Hubbard model to describe the overdoped region, Ossadnik et al. [\onlinecite{ossadnik2008}] found good qualitative agreement with the experimental results on overdoped Tl2201 for the temperature and angular dependence of the relaxation rate determined from the magnetoresistance [\onlinecite{abdel2006}]. Since overdoped Tl2201 shows a conventional full Fermi surface in angle-resolved photoemission spectroscopy (ARPES) at zero magnetic field [\onlinecite{TI2201FullFermiSurfaceARPES}] and in quantum oscillation experiments at finite magnetic fields [\onlinecite{TI2201FullFermiSurfaceQuOsc}] a perturbative treatment is of interest. Note that the  FRG calculations are evaluated only to one-loop order and therefore are quantitatively reliable only for weak to moderate values of the Hubbard onsite interaction, $U$. As a result, a quantitative comparison to the experimental data shows a substantial underestimate of the magnitude of the calculated temperature dependent resistance, e.g. see the comparison in Kokalj and McKenzie [\onlinecite{kokalj2011}].

In their calculation of unusual transport behavior, Ossadnik and collaborators analyzed the quasiparticle scattering vertex for overdoped cuprates within a functional renormalization group (FRG) approach [\onlinecite{ossadnik2008}]. They found  an anisotropic term in the quasiparticle scattering rate with an angular form similar to the experiment and with an unconventional $T$-linear dependence which increased when approaching optimal doping starting  from the overdoped side. They argued  that this unusual term appearing in the imaginary part in the quasiparticle self-energy could lead also to an equivalent temperature dependence for the in-plane resistivity.

The goal of this study is to expand this idea and move beyond a self energy analysis of the transport properties to a full calculation of transport properties using solutions of the Boltzmann equation. This is possible within a FRG approach since the method  computes the strongly renormalized quasiparticle interactions. We determine the matrix elements for the scattering events within the FRG framework presented in [\onlinecite{ossadnik2008}] and use numerical techniques to solve the linearized Boltzmann equation. A detailed numerical treatment is required in view of the strong anisotropy in the transport life times [\onlinecite{chang2008,yoshida2007}] due to both the anisotropic quasiparticle interactions and the strongly angle dependent Fermi velocities introduced by the anisotropy of the band structure. The importance of full transport calculations was stressed by Hlubina and Rice in an early paper [\onlinecite{hlubina1995}].    

As will be discussed later, our calculations agree qualitatively with the unconventional temperature dependence but  the magnitude of the resistance is substantially lower. This disagreement in the absolute values is in line with the restrictions on $U$ required to justify the one-loop approximation in the FRG. Another discrepancy between our numerical results and the experiment is the lack of scaling of the resistance coefficients, $A_{1\square}$ and $A_{2\square}$, with $1/p$ the inverse hole density as reported by Bari\v{s}i\'{c} et al. [\onlinecite{greven2012}]. Note however, our results are for the overdoped samples. The data in Fig. 5 of Ref. [\onlinecite{greven2012}] deviate from the inverse doping scaling in the overdoping density range. In the case of the linear coefficient, $A_{1\square}$ the data show a linear decrease and vanish at the onset of superconductivity in line with the earlier results of Hussey and coworkers [\onlinecite{hussey2009}].

A good example of such an onset of a non-Fermi-liquid-like $T$-linear contribution to the resistivity can be found in the resistivity measurements by Hussey and coworkers [\onlinecite{hussey2009}] upon reducing the hole concentration from the overdoped regime toward optimal doping in \lsco\ (cf. Fig. \ref{fig: experimental resistivity}). They observed a strong growth of the temperature dependent contribution to the resistivity with a decrease in the hole doping together with a remarkable trend toward linear resistivity that goes down to lowest temperatures. They used strong magnetic fields to explore the normal state transport properties within the superconducting dome. The analysis of the temperature dependence of the resistivity was based on a second-order polynomial fit to the experimental data over a wide temperature range: $ \rho(T) = \alpha_0 + \alpha_1 T + \alpha_2 T^2 $. This allows to distinguish doping regimes with dominant linear versus dominant quadratic temperature dependence.

We also solve the Boltzmann equation to obtain the Seebeck coefficient which describes the charge transport in response to a thermal gradient. Standard Fermi liquid theory predicts a metallic Seebeck coefficient with a linear $T$-dependence.  The experiments by Lalibert\'{e} and coworkers [\onlinecite{laliberte2011}] showed substantial deviations from the standard form with a complex density dependence. Again our calculations reproduce the qualitative trends in the temperature and density dependence very well.

In Sec. \ref{sec: Numerical Results} we show our numerical results and compare with the experimental data of Refs. [\onlinecite{hussey2009,hussey2011,greven2012}]. We also examine the validity of Matthiessen's rule which relies on a decoupling of impurity and two-particle scattering in Sec. \ref{sec: Momentum relaxation mechanism in the cuprates and Matthiessens rule}. Finally in Sec. \ref{sec: the seebeck effect in overdoped lsco}, we study the thermoelectric effect of \lsco, use the Seebeck coefficient as a sign of non Fermi liquid transport and examine the evidence for quantum critical behavior.

The numerical method that we have developed to study the transport properties of correlated materials such as \lsco\ is explained in detail in the appendix, App. \ref{sec: model for the quasiparticle dispersion of lsco}-\ref{sec: Effective Model for the Normal State Charge Transport of the High-Temperature Superconductor LSCO}. We solve the semiclassical Boltzmann equation taking the full angular and energy dependence of the distribution function into account. The two important inputs of the model are the explicit form of the band structure and the temperature dependent calculation of the quasiparticle scattering rates.

\section{Numerical Results}
\label{sec: Numerical Results}

We begin by reviewing our numerical results. We use a semiclassical model to describe the normal state charge transport of overdoped \lsco. For the quasiparticle dispersion we used a two-dimensional tight binding model and a doping dependent hopping parameter renormalization scheme based on the values determined in Ref. [\onlinecite{yoshida2006}] from the Fermi surface in APRES data on \lsco. The use of phenomenological input for the quasiparticle dispersion is important because the renormalization of the hopping integrals is of two-loop order in the FRG analysis that we use to determine the scattering vertex. The calculation of two-loop diagrams for the RG flow is, however, beyond the scope of this study. For details about the dispersion, cf. App. \ref{sec: model for the quasiparticle dispersion of lsco}. The FRG technique that we use to compute the renormalized quasiparticle scattering rates is outlined in App. \ref{sec: Strongly renormalized quasiparticle interactions} together with brief overview of the doping and temperature dependence of the scattering vertex and a discussion of the parameters that enter the RG calculation. 

Taking the quasiparticle dispersion and the scattering vertex as input, the conductivity is computed by solving the linearized Boltzmann equation with its full angular and energy dependence. Due to the strong anisotropy in the scattering rates and the quasiparticle velocities, a full solution of the Boltzmann equation rather than a single relaxation time approximation is required. The collision integral is computed by introducing an efficient discretization of the Brillouin zone and the Boltzmann equation is then solved numerically. This method of solving the Boltzmann equation and discretizing the Brillouin zone is explained in App. \ref{sec: Effective Model for the Normal State Charge Transport of the High-Temperature Superconductor LSCO}.

\subsection{Simulation estimates of $\rho(T)$ for \lsco\ in a range of doping}
\label{subsec: Simulation of the rho for lsco with variable doping}

\begin{figure*}[t]
\includegraphics[width=0.99\textwidth]{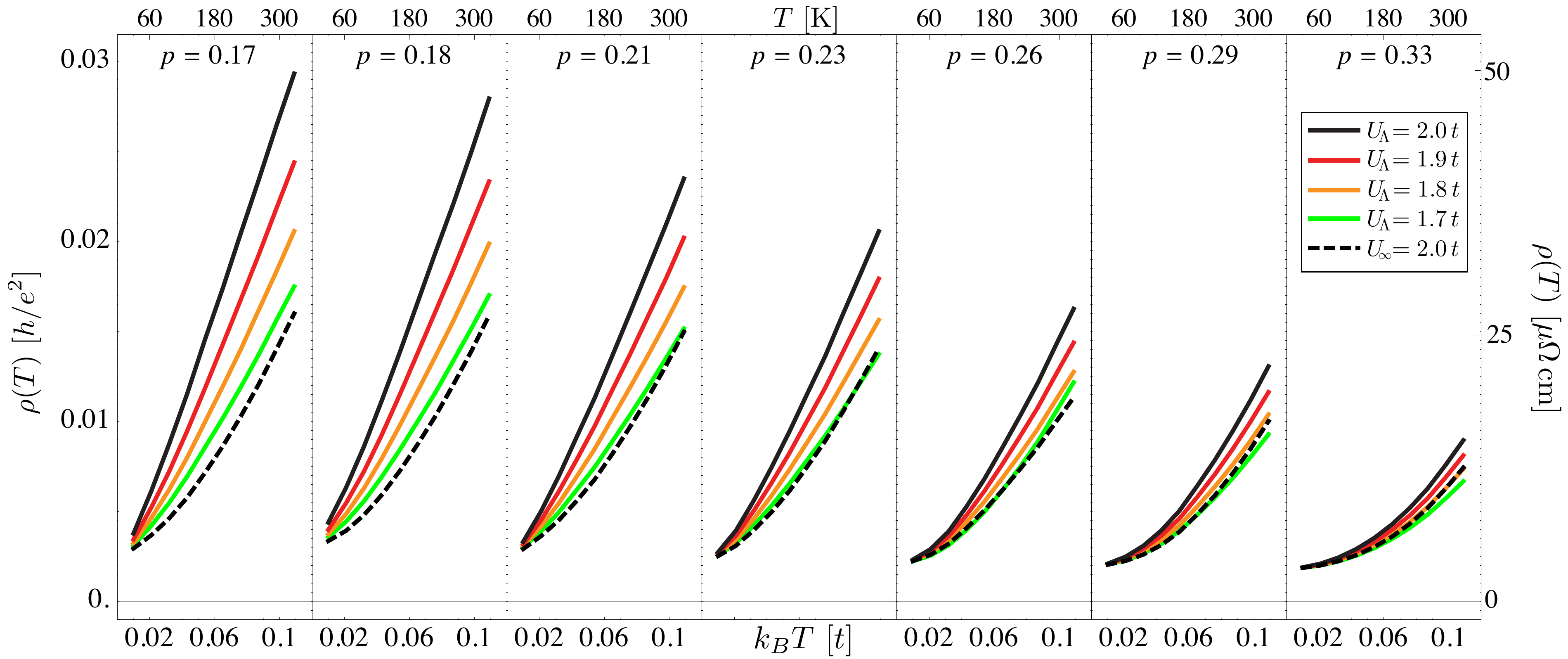}
\caption{Numerical calculation of the normal state resistivity of \lsco\ (black line, $U_{\Lambda}=2.0\,t$) as a function of the temperature for several doping levels: The resistivity is plotted for several values for the initial onsite repulsion $U$. The RG cutoff is given by $\Lambda=0.01\,t$, except for the dashed line ($U_{\infty}$) computed from unrenormalized scattering rates. The $T\rightarrow 0$ extrapolation of $\rho$ corresponds to the residual resistivity due to impurity scattering ($n_{\text{imp}}W^{2}_{\delta\text{-imp}}=0.25\,t^{2}$). We have converted the temperature and resistivity to the units used in the experiment. For details cf. App. \ref{sec: conversion of computation to experimental units}.}

\label{fig: rhoVsTemperatureLSCO}

\end{figure*}

The in-plane resistivity in our model of the normal state of overdoped \lsco\ is shown in Fig. \ref{fig: rhoVsTemperatureLSCO} (data for $U_{\Lambda}=2.0\,t$). We also plot the resistivity for different initial values for the unrenormalized onsite interaction parameter $U$. The data corresponding to the different values of $U_{\Lambda}$ are derived from renormalized scattering vertices while the data for $U_{\infty}$ are computed for particles interacting via the bare onsite repulsion. For a comparison of these calculations, cf. Sec. \ref{subsec: discussion of the parameters}. The values for the hole-doping $p$ are chosen to match the experimental measurements of Fig. \ref{fig: experimental resistivity}.

Our results compare well to many qualitative aspects of the experimental data of Ref. [\onlinecite{hussey2009}]. First, there is the strong growth of the resistivity when reducing the hole doping from strongly overdoped values of $p\approx 0.35$ to optimal doping around $p\approx 0.17$. Second, there is clear evidence for a growing linear term in the temperature dependence of $\rho$ around optimal doping. For large values of the hole doping, $\rho(T)$ displays a rather quadratic temperature dependence. On the other hand, for doping concentrations below $p\approx 0.25$ a linear $ T $-dependence becomes obvious. We analyze our data in the same scheme as in the experimental paper [\onlinecite{hussey2009}] using fits with the form $\rho(T)=\alpha_{0}+\alpha_{1}T+\alpha_{2}T^{2}$ (Sec. \ref{subsubsec: linear plus quadratic fit}) or also by introducing a "local scaling exponent" $x(T,p)$, e.g. $\rho(T)=\alpha_{0}+\alpha_{\infty}T^{x}$ (Sec. \ref{subsubsec: local scaling exponent}).

For a quantitative comparison to the experiment we convert our two-dimensional data to the experimental units of a three-dimensional sample (cf. App. \ref{sec: conversion of computation to experimental units}). From Fig. \ref{fig: rhoVsTemperatureLSCO} we find that our calculated values for $\rho(T)$ at $T\approx 200\,\textrm{K}$ are roughly by a factor four or five smaller than in experiment (cf. Fig. \ref{fig: experimental resistivity}). An underestimation of the resistivity is consistent with the fact that our unrenormalized interaction strength $ U = 2.0\,t$ is considerably lower than expected for \lsco.

In spite of the weak coupling approximation in our calculation of the scattering rates, the resistivity is in good qualitative agreement with the observed temperature dependence.

\subsection{Scaling behavior}
\label{subsec: Scaling behavior}

\subsubsection{$T+T^{2}$ fit to the resistivity curves}
\label{subsubsec: linear plus quadratic fit}

\begin{figure*}[t]

\includegraphics[width=0.45\textwidth]{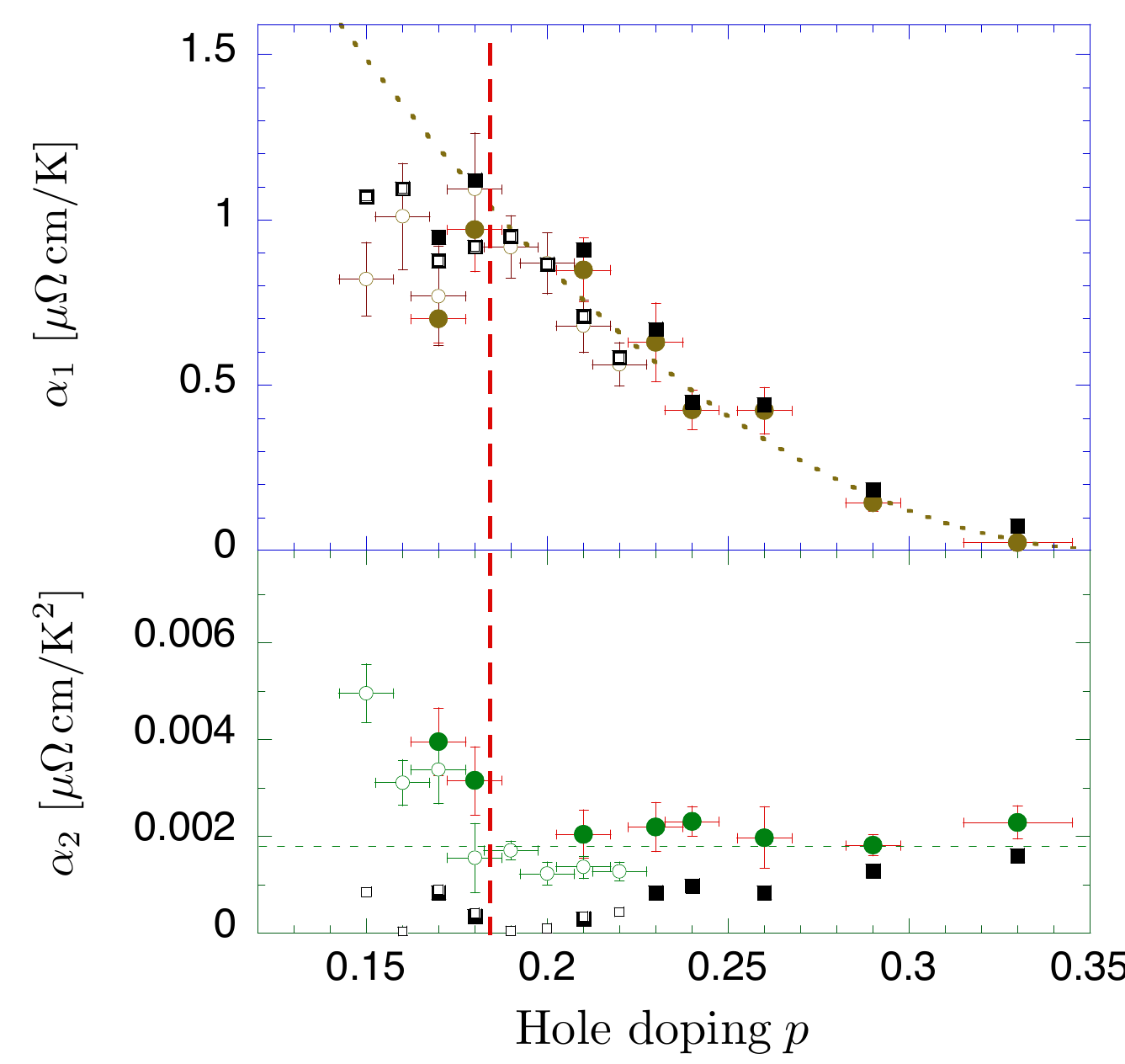}\qquad
\includegraphics[width=0.45\textwidth]{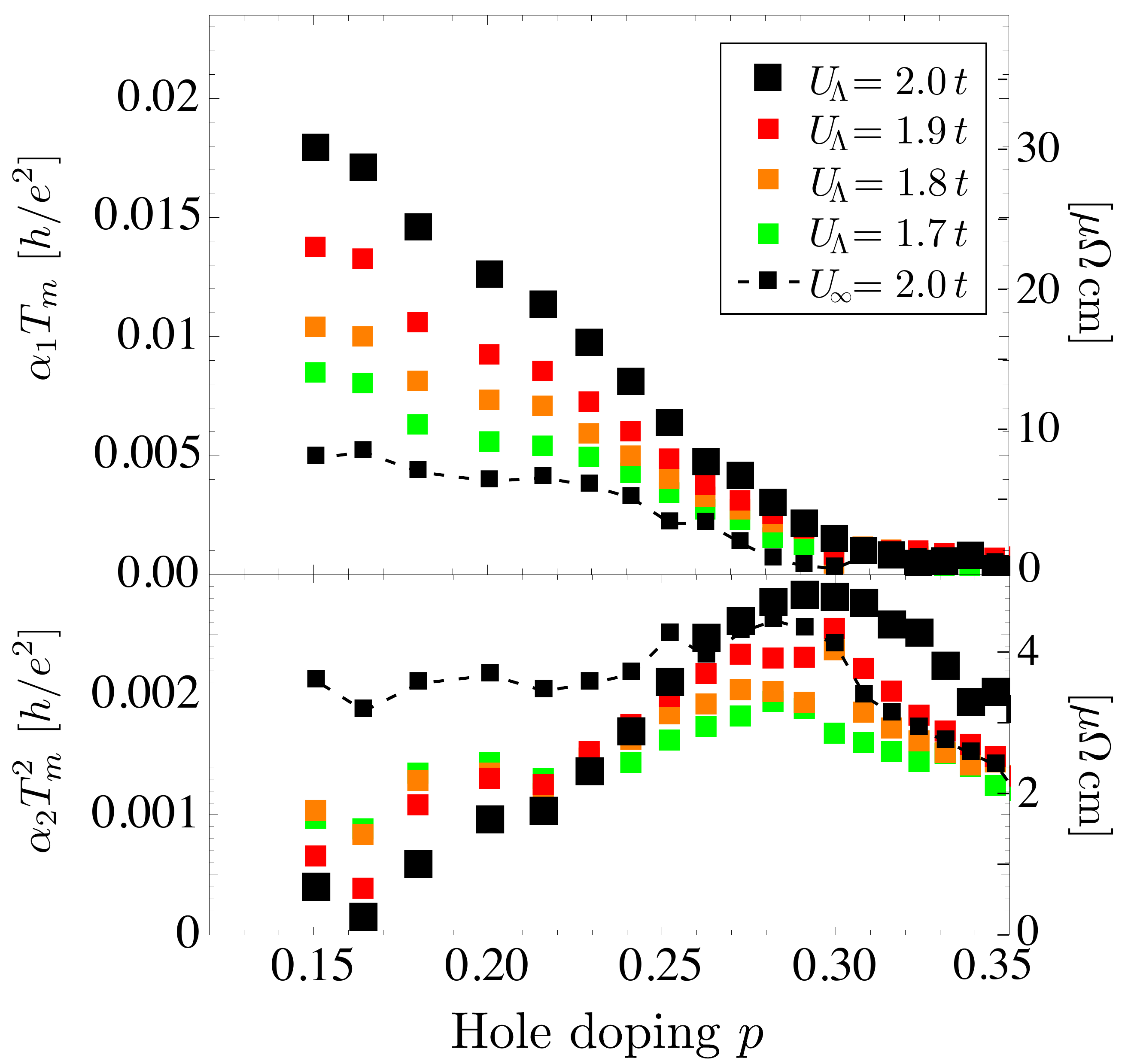}
\caption{\emph{left}: Experimental results obtained by Cooper et al. [\onlinecite{hussey2009}]. The quantities $\alpha_{1}$ and $\alpha_{2}$ are coefficients of fit to the experimentally measured resistivity (Fig. \ref{fig: experimental resistivity}): black filled squares: fit as in our numerical analysis, $\rho=\alpha_{0}+\alpha_{1}T+\alpha_{2}T^{2}$-fit; green filled circles: parallel resistor fit, $1/\rho=1/(\alpha_{0}+\alpha_{1}T+\alpha_{2}T^{2}) + 1/\rho_{\text{max}}$; open symbols correspond to the same analysis for experimental data of Ando et al. [\onlinecite{ando2004}] \emph{right:} Linear ($\alpha_{1}$) and quadratic ($\alpha_{2}$) term of the $\rho=\alpha_{0}+\alpha_{1}T+\alpha_{2}T^{2}$ fit to our numerical resistivity data of Fig. \ref{fig: rhoVsTemperatureLSCO} as a function of the hole-concentration $p$: Shown are the coefficients for different initial values of the onsite repulsion $U$ and for the calculation with unrenormalized interactions ($U_{\infty}$). We convert the coefficients to resistivity units by introducing the average temperature scale $T_{m}=150\,\textrm{K}$ of the fit regime.} 
\label{fig: coefficientsLSCO}
\end{figure*}

Next we analyze the numerical resistivity data using the polynomial fit up to second order $\rho(T)=\alpha_{0}+\alpha_{1}T+\alpha_{2}T^{2}$, in analogy to Cooper et al, cf. Ref. [\onlinecite{hussey2009}]. They fit their resistivity data over the entire temperature range and obtain the fit coefficients $\alpha_{1}$ and $\alpha_{2}$ as a function of the hole doping, cf. Fig. \ref{fig: coefficientsLSCO} (left panel). They found the linear coefficient $\alpha_{1}$ grows strongly with a decrease in doping while there is a constant or decreasing value of $\alpha_{2}$. 

The linear $\alpha_{1}$ and quadratic $\alpha_{2}$ coefficients from fitting our numerical results are shown in Fig. \ref{fig: coefficientsLSCO} (right panel) for the entire range of hole doping from $p\approx 0.15$ to $p\approx 0.35$. The fit was performed for temperatures ranging from $30\,\textrm{K}$ up to $300\,\textrm{K}$. For the comparison of the two coefficients we introduce an average temperature scale $T_{m}=150\,\textrm{K}$ in the conversion to resistivity units.  The coefficient representing the residual resistivity, $\alpha_{0}$, is discussed later.

The direct comparison of our fit coefficients to the experiment yields very good qualitative agreement. The linear term shows a strong increase with reduced hole concentration toward optimal doping while above $p \approx 0.30$ it becomes indistinguishable from zero. On the other hand, the quadratic contribution does not grow and even decreases almost down to zero toward optimal doping for larger $U$. A vanishing $\alpha_{2}$ indicates a within our model a purely linear temperature dependence. Whether or not the quadratic coefficient of the fit to the experimental data vanishes is not really clear from Ref. [\onlinecite{hussey2009}], although it appears that the $\alpha_{2}$ determined with the conventional $T+T^{2}$ fit indeed decreases and finally vanishes around $p\approx 0.19$, while the quadratic coefficient determined from a parallel resistor fit remains constant.

The coefficients in the overdoped regime disagree with the reported $1/p$ dependence by Bari\v{s}i\'{c} et al. [\onlinecite{greven2012}]. As already mentioned in the introduction, in the overdoped regime, their data suggest a deviation from the remarkable $1/p$ dependence of the linear term in the underdoped regime. The experimental data of both Ref. [\onlinecite{hussey2009}] and [\onlinecite{greven2012}] indicate that the linear term eventually goes to zero for sufficiently large hole density, $p\approx 0.30$.

\subsubsection{Scaling regimes}
\label{subsubsec: scaling regimes}

\begin{figure}[t]

\includegraphics[width=0.33\textwidth]{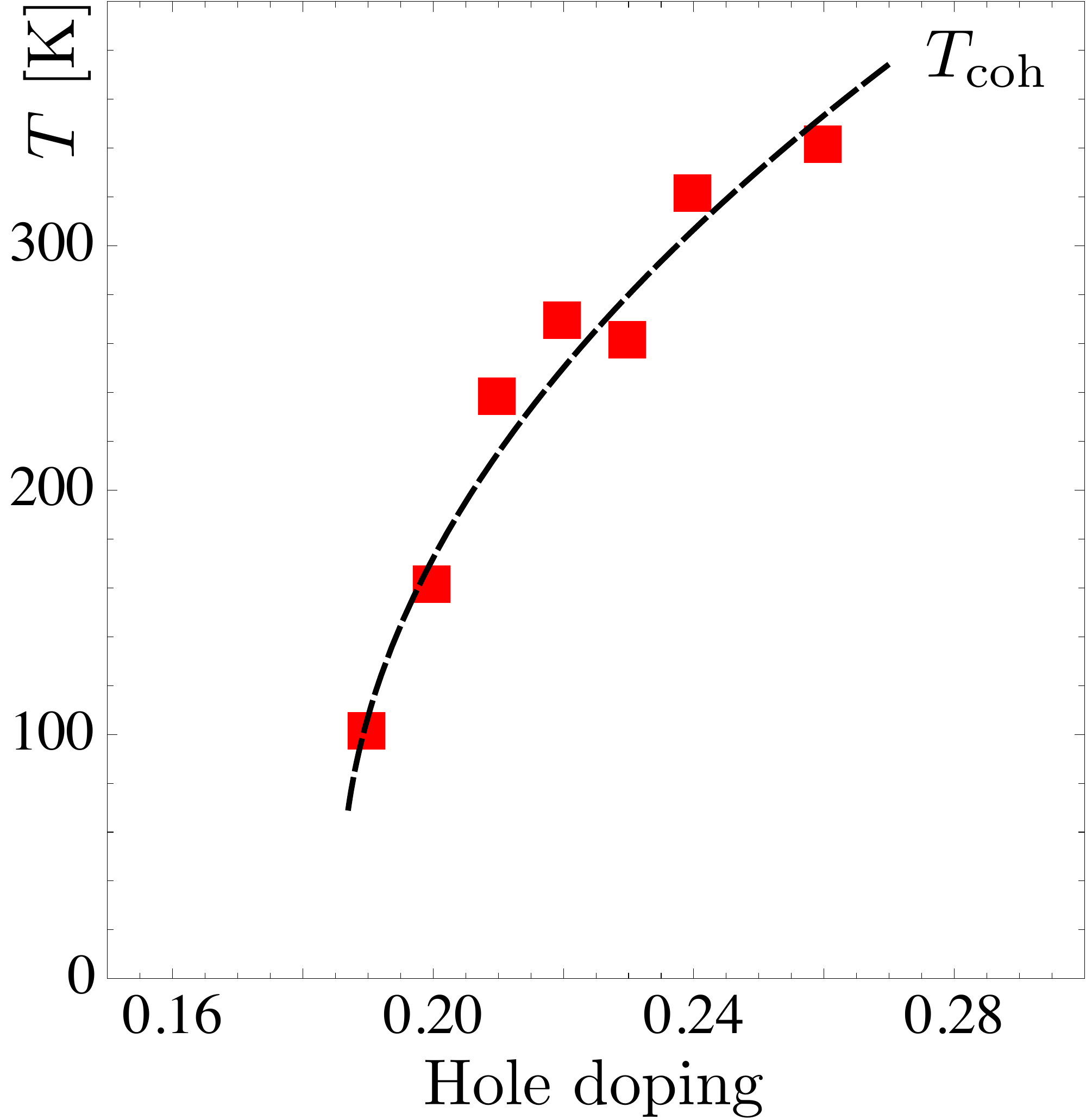}

\caption{
Experimentally determined coherence temperature $T_{\text{coh}}$ ($p>0.19$) as a function of the hole doping of \lsco. The coherence temperature is where the temperature derivative of the conductivity deviates from its constant value at high temperature (data adopted from Ref. [\onlinecite{hussey2011}]).
}
\label{fig: phase diagram and coherence temperature} 

\end{figure}

An alternative analysis of the experiments was recently put forward by Hussey et al. [\onlinecite{hussey2011}]. Taking the derivative $\rho'=d\rho(T)/dT$ they distinguished two regimes which are not immediately obvious in the resistivity data, but in which $\rho(T)$ scales differently with temperature. Hussey and coworkers identified a high-temperature regime where $\rho'$ is constant, i.e. the resistivity depends linearly on $ T $. Toward lower temperatures $ \rho' $ deviates downward from this constant value. With this rather abrupt change one can associate a temperature $T_{\text{coh}}$, called ''coherence temperature'' in Ref. [\onlinecite{hussey2011}] which decreases with lowering the hole concentration and disappears probably around the same value of $p$ where the characteristic temperature for the pseudogap vanishes extrapolated from the underdoped side. In Fig. \ref{fig: phase diagram and coherence temperature} the doping dependence of $T_{\text{coh}}$ measured in Ref. [\onlinecite{hussey2011}] is shown. It was experimentally shown that $T_{\text{coh}}$ coincides with the loss of the quasiparticle coherence peak [\onlinecite{kaminski2003}].

\begin{figure*}[t]

\includegraphics[width=0.48\textwidth]{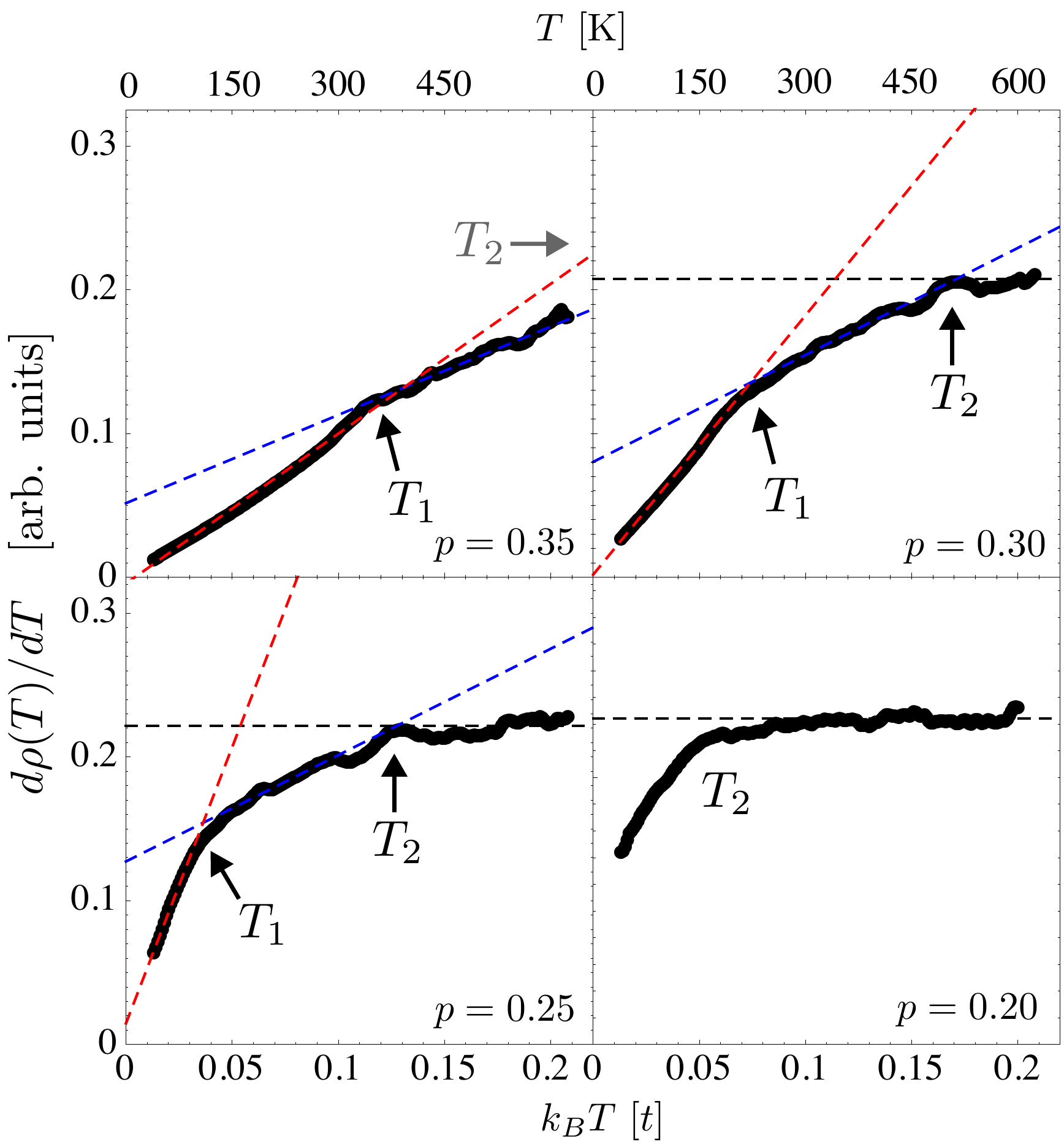}\qquad
\includegraphics[width=0.48\textwidth]{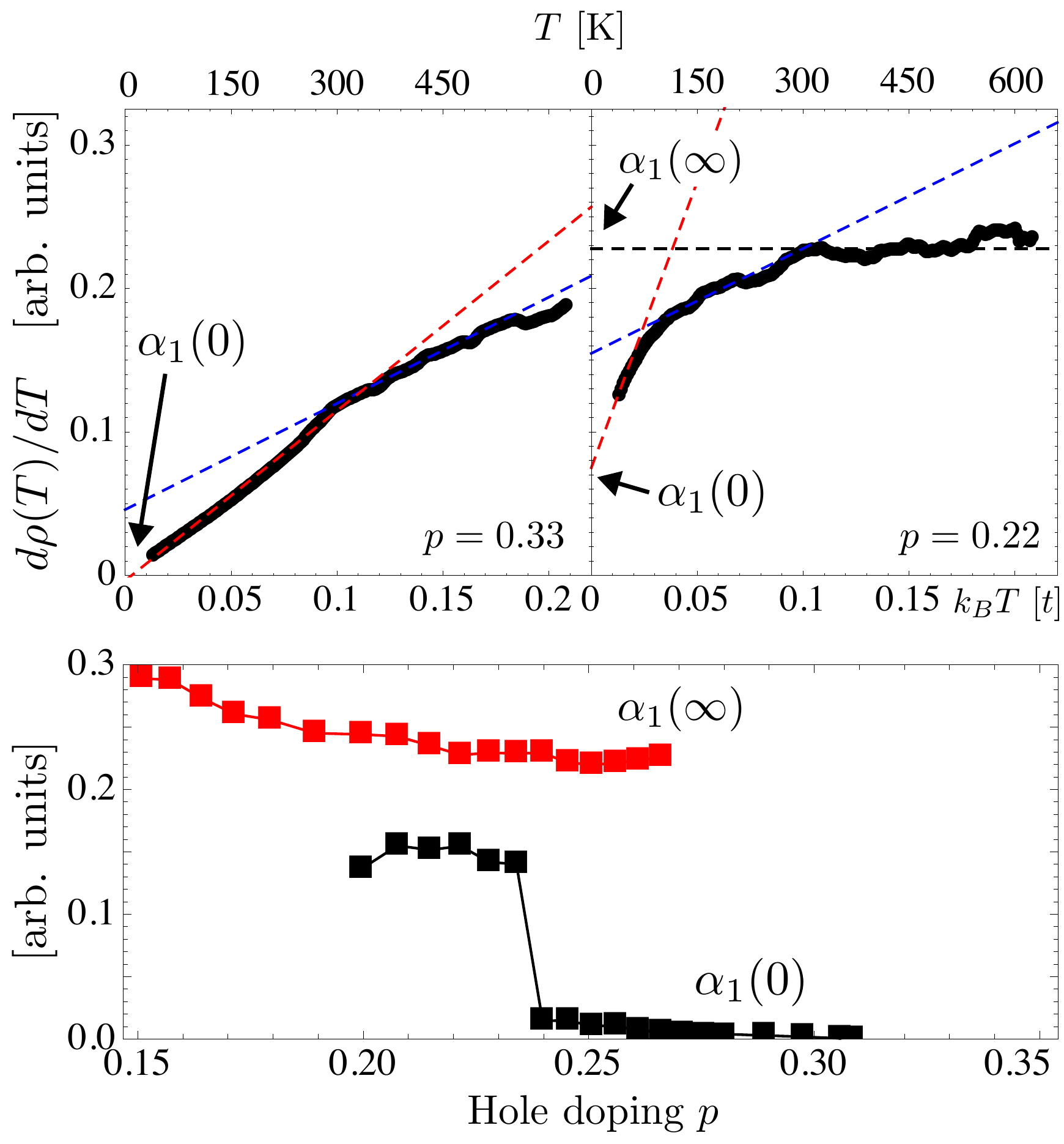}
\caption{\emph{left}: Plot of the temperature derivative of the resistivity $\rho'(T)$: Three intervals are identified in which $\rho(T)$ shows different scaling behavior. Shown is $\rho'(T)$ for four different values of the hole doping illustrating the doping dependence of $T_{1}$ and $T_{2}$. For the bottom right inset at doping level $p\approx0.20$, the temperature scales cannot be clearly identified any more. In Fig. \ref{fig: ExponentPhaseDiagram} $T_{1}$ and $T_{2}$ are plotted as a function of the doping. 
\emph{right}: Limiting behavior of the linear term that dominates the temperature dependence of the resistivity: The coefficient $\alpha_{1}(0)$ represents the $T\rightarrow 0$ extrapolation of the slope in the resistivity curve while $\alpha_{1}(\infty)$ corresponds to the coefficient of the dominating linear term in the high-temperature regime. The strong growth of $\alpha_{1}(0)$ for doping levels $p<0.25$ reflects the fact that the unconventional scaling behavior extends down to low temperature for these dopings. The definition of $\alpha_{1}(0,\infty)$ only allows these quantities to be obtained in a finite doping range.}
\label{fig: coherence_temperature}

\end{figure*}

Plotting $\rho'(T)$ for different doping levels in our calculation, we can analogously identify distinct regimes in the temperature scaling of the resistivity, cf. Fig. \ref{fig: coherence_temperature} (left panel). There is a low-temperature regime with finite slope of $\rho'(T)$, an intermediate temperature regime with finite, but smaller slope, and a high-temperature regime where $\rho'(T)$ is constant. The temperatures that separate these three regimes will be referred to as $T_{1}$ and $T_{2}$. 

The temperature scales $T_{1}$ and $T_{2}$ are displayed in Fig. \ref{fig: ExponentPhaseDiagram}. The doping dependence of $T_{2}$ shows a very similar trend as $T_{\text{coh}}$. This is interesting because there is no real loss of quasiparticle coherence in our semiclassical model based on the Boltzmann transport equation. Note that we have not considered the values for $T_{1}$ and $T_{2}$ for $p<0.2$ as it becomes difficult to distinguish the regimes for $p\approx 0.19$

It is interesting to relate the onset of the unconventional scaling, $T_{1}$, to the proximity of the Fermi energy to the van Hove level, $E_{\text{vH}}$. We introduce the temperature scale $T_{\Delta_{\text{vH}}} = 4|\mu-E_{\text{vH}}|/k_{B}$, which is shown as the red squares in Fig. \ref{fig: ExponentPhaseDiagram}. The factor of 4 is introduced for convenience, as we consider only those low-energy states relevant for the transport, that live within an energy range of $4k_{B}T$ from the Fermi level. The doping dependence of $T_{\Delta_{\text{vH}}}$ and $T_{1}$ is very similar and we conclude that proximity to a van Hove singularity plays an important role in the development of non-Fermi liquid behavior.

To compare to the analysis in Ref. [\onlinecite{hussey2011}] we fit the high-temperature linear part of $\rho'$ with a constant $\alpha_{1}(\infty)$ and the low-temperature regime by a linear function $\alpha_{1}(0)+\alpha_{2}T$. In the high-temperature regime, $\alpha_{1}(\infty)$ represents the coefficient of the dominating linear term in the resistivity, while $\alpha_{1}(0)$ corresponds to the coefficient of the dominating linear term in the limit $T\rightarrow 0$. The results for these coefficients are shown in Fig. \ref{fig: coherence_temperature} (right panel).

\subsubsection{Derivation of the local scaling exponent}
\label{subsubsec: local scaling exponent}

\begin{figure}[t]
\begin{center} 

\includegraphics[width=\columnwidth]{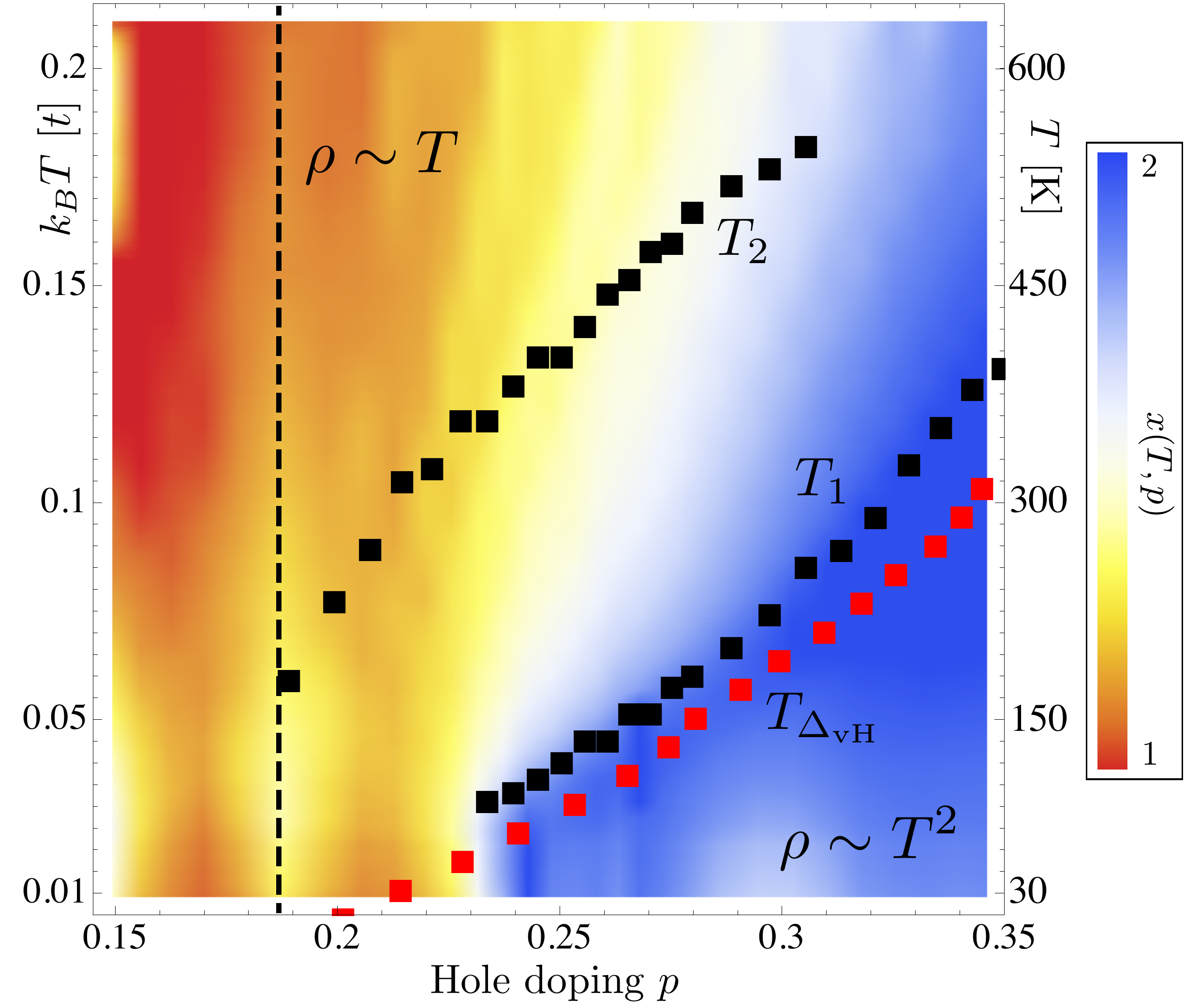}

\caption{Temperature and doping dependence of the scaling exponent $x(T,p)$ defined in Eqs. \eqref{eq: definition of the local exponent} and \eqref{eq: logarithmic derivative of rho yields x} for our model of \lsco\ in color-code. The temperature scales $T_{1}$ and $T_{2}$, cf. Sec. \ref{subsubsec: scaling regimes}, are shown as black squares. The temperature scale $T_{\Delta_{\text{vH}}}$ corresponds to the energy difference between the chemical potential and the energy level of the van Hove singularity, $k_{B}T_{\Delta_{\text{vH}}} = 4|\mu-E_{\text{vH}}|$ (we consider low-energy states up to an energy of $4k_{B}T$ around the Fermi energy). Note that the unconventional scaling $x<2$ goes down to the lowest temperatures at optimal doping. The Lifshitz transition is indicated as the horizontal dashed line.}
\label{fig: ExponentPhaseDiagram} 

\end{center}
\end{figure}

A different way to illustrate the regime change on the resistivity data is to define an exponent of $\rho(T)$ for each temperature formulated as
\begin{align}
	\rho(T) = \alpha_0 + \alpha T^{x},\qquad x=x(T).
\label{eq: definition of the local exponent}
\end{align}
The exponent $x(T)$ is obtained by taking the logarithmic derivative of the $\rho(T)$,
\begin{align}
	x(T) = \frac{\dd}{\dd \ln T} \ln(\rho(T) - \alpha_{0}),
\label{eq: logarithmic derivative of rho yields x}
\end{align}
where $\alpha_{0}$ denotes the residual resistivity. The results for $x(T)$ over a wide range in doping are shown in Fig. \ref{fig: ExponentPhaseDiagram}.
 
Even though such a local scaling law has large uncertainties, it is intriguing to find a very good qualitative match between the function of $x(T,p)$ and the generic scaling behavior of the cuprates, cf.  and Ref. [\onlinecite{hussey2008}]. In the overdoped low-temperature region of the phase diagram, the local exponent rises to the value of $x\approx 2$ corresponding to a standard Fermi liquid. Around optimal doping, $x(T,p)$ is always well below $2$ down to the lowest temperature of our study. In the high-temperature regime, the onset of unconventional scaling ($x(T,p)<2$) is shifted to higher temperature with increasing doping.

\subsubsection{Angular resolved scaling analysis}
\label{subsubsec: Angular resolved scaling analysis}

\begin{figure}[t]
\begin{center}

\includegraphics[width=0.9\columnwidth]{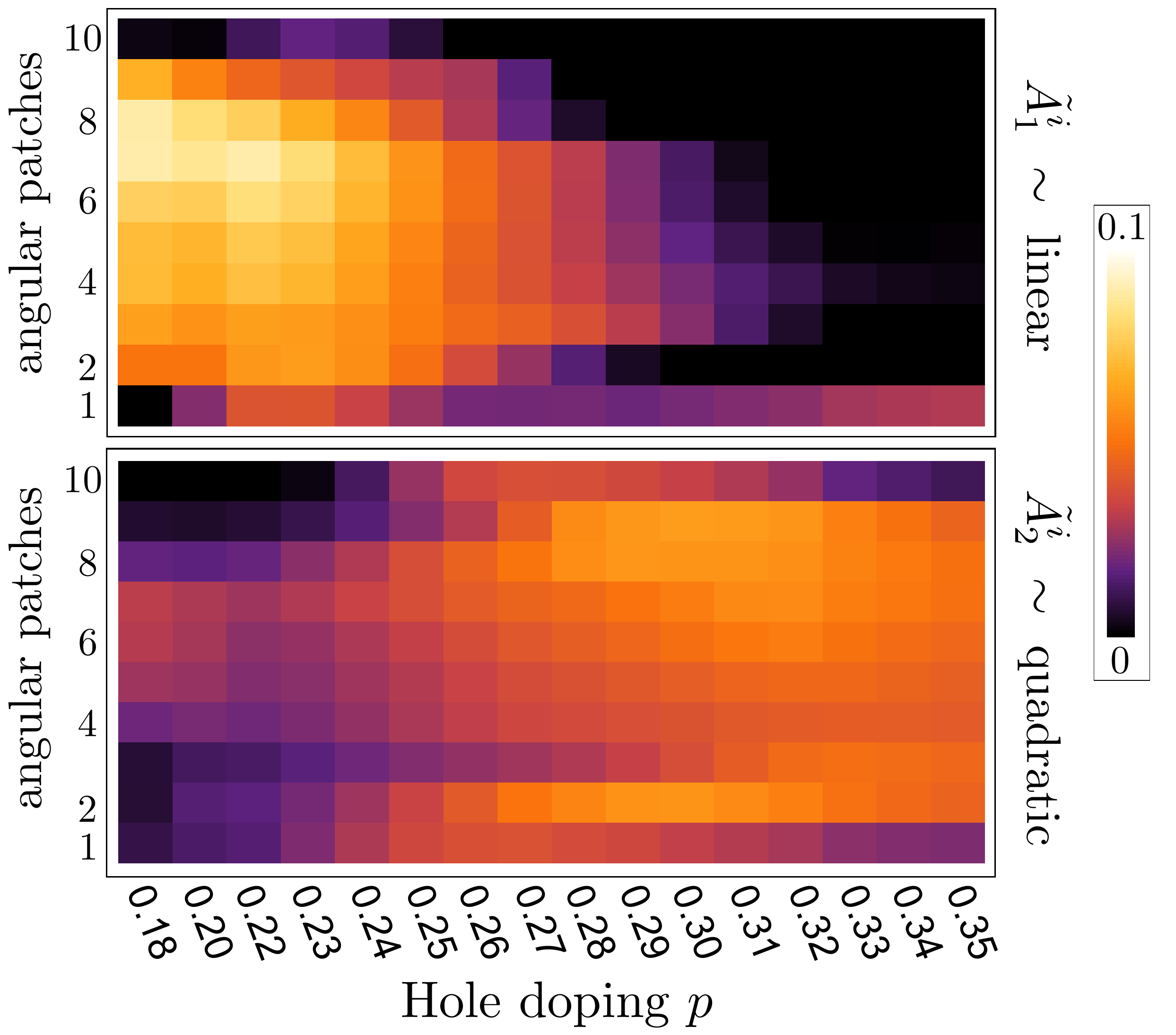}
\caption{Comparison of linear vs. quadratic temperature scaling of the resistivity for the angular patches (cf. Fig. \ref{fig: discretization}). On top, we have plotted the weighted linear term $\tilde{A}^{i}_{1}$, and the lower plot shows the weighted quadratic term $\tilde{A}^{i}_{2}$, defined in Eq. \eqref{eq: definition of As}. These $\tilde{A}^{i}_{a}$'s are weighted according to the contribution of the angular patch $i$ to the total charge transport. The patches range in angular direction from $\theta=0$ (patch 1) to $\theta=\pi/2$ (patch 10). The $\tilde{A}^{i}_{a}$'s are evaluated at the temperature $T_{m}=150\,\textrm{K}$, the average temperature of the fit region.}
\label{fig: angularResolutionCoefficients}

\end{center}
\end{figure}

Using the non-equilibrium distribution function, we can analyze the angular dependence of this unconventional scaling and compare to the difference in the temperature dependence between isotropic and anisotropic scattering rates in the FRG study [\onlinecite{ossadnik2008}]. The quasiparticle scattering rates are strongly anisotropic and especially the quasiparticles close to the saddle points (anti-nodal-direction) in the Brillouin zone have very short transport life times. 
The scattering rates that cause these short life times are strongly enhanced as the doping level decreases toward optimal doping. To examine the role of strong scattering of the anti-nodal quasiparticles for the unconventional temperature dependence of the overall resistivity, we interpret the angular patches of our Brillouin zone discretization (cf. Fig. \ref{fig: discretization} in the appendix) as individual resistors and parametrize the low-energy quasiparticle states as a parallel resistor network. Each resistor is characterized by the conductivity $\tilde{\sigma}_{i}$ and resistivity $\tilde{\rho}_{i} = 1/ \tilde{\sigma}_{i} $ where the index $i$ labels the the angular patch. The total conductivity and resistivity are determined by
\begin{align}
	\sigma = \sum_{i} \tilde{\sigma}_{i}, \qquad \rho=\biggl(\sum_{i}1/\tilde{\rho}_{i}\biggr)^{-1}.
\label{eq: conductivity and resistivity separated into angular pieces}
\end{align}
In Fig. \ref{fig: angularResolutionCoefficients} we depict the the angle-resolved scaling by plotting the linear and quadratic contributions to the patch resistivity as functions of the angular patch index $i$ and the hole concentration. The temperature dependence of each resistor was analyzed by a linear plus quadratic fit,  $\tilde{\rho}_{i}(T)=\tilde{\alpha}_{0}^{i}+\tilde{\alpha}_{1}^{i}T+\tilde{\alpha}_{2}^{i}T^{2}$. In order to compare linear and quadratic term, we give them a weight factor that represents their contribution to the total resistivity and normalize their sum, the total resistivity $\rho$, to unity,
\begin{align}
	1&=
	\sum_{i}\frac{\rho}{\tilde{\rho}_{i}}\frac{\tilde{\alpha}_{0}^{i}+\tilde{\alpha}_{1}^{i}T+\tilde{\alpha}_{2}^{i}T^{2}}{\tilde{\rho}_{i}}= \sum_{i}\tilde{A}^{i}_{0}+\tilde{A}^{i}_{1}+\tilde{A}^{i}_{2},
\label{eq: weight factor angular coefficients}
\intertext{with}
	&\qquad\tilde{A}^{i}_{a}=\frac{\rho}{\tilde{\rho}_{i}^{2}} \,\tilde{\alpha}^{i}_{a}\,T^{a},\quad a=0,1,2.
\label{eq: definition of As}
\end{align}
Thus $\tilde{A}^{i}_{a}$ represents the normalized contribution of the angular patch $i$ to the term in the total resistivity that scales with temperature exponent $a$.

From Fig. \ref{fig: angularResolutionCoefficients} one can clearly see that in the strongly overdoped regime only the quadratic term is present and is essentially independent of the angle. On the other hand,  below $p\approx0.30$, the linear term grows showing strong anisotropy. The linear term appears first close to the nodal direction (patches 4 to 7), and constantly spreads out. Below $p\approx0.27$ the linear term starts to dominate the transport. The quadratic term, however, remains finite in the nodal direction down to the lowest doping levels.

It is interesting to see that around optimal doping, where the Fermi surface is very close to the van Hove singularity, the charge transport from the angular patches 1 and 10, that contain the saddle points, essentially vanishes. This is consistent with the strong growth of the scattering rates from the FRG analysis. These angular patches are short-circuited as a consequence of strong scattering, cf. Ref. [\onlinecite{hlubina1995}]. Our discussion here is fully compatible with the behavior of the angle-dependent scattering rates found by the FRG study by Ossadnik and collaborators [\onlinecite{ossadnik2008}].

\begin{figure*}[t]
\begin{center}

\includegraphics[width=\textwidth]{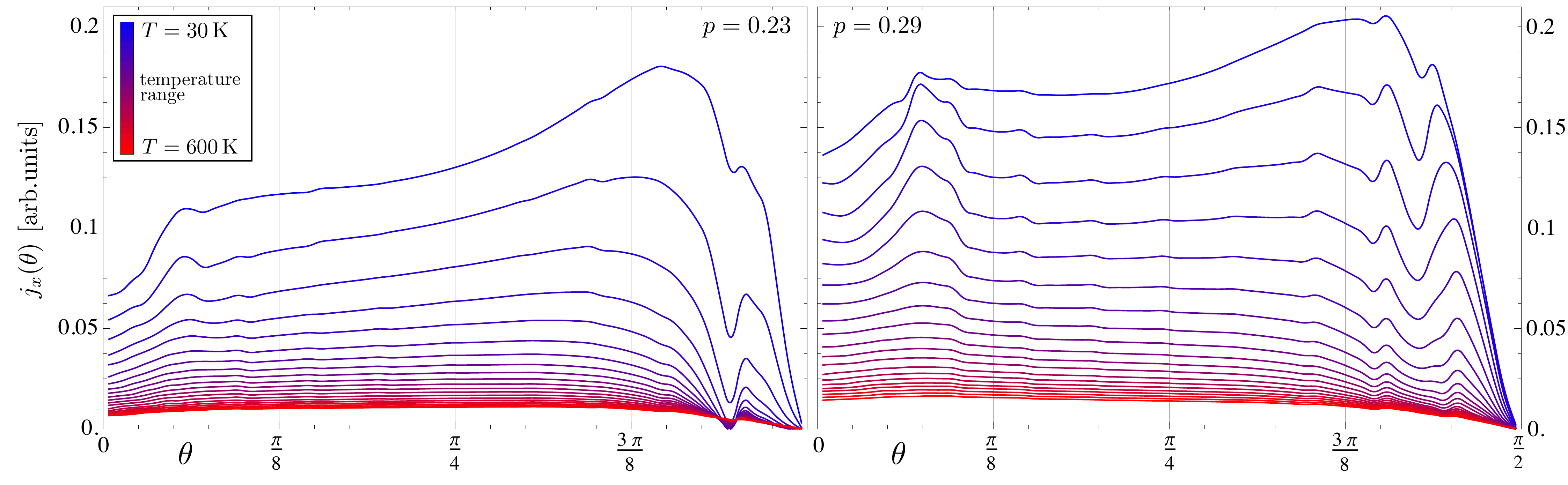}
\caption{Plot of the angular distribution of the current for doping levels $p=0.23$ and $p=0.29$. The curves correspond to temperatures between $30$ and $600\,\textrm{K}$ indicated by their color. Shown is merely the angular range between $0$ and $\pi/2$ due to the four-fold lattice symmetry. Note that the dip between $3/8\,\pi$ and $1/4\,\pi$ comes from the $(2\pi/a,0)$-umklapp scattering process. There are many states close to $(\pi-\delta,0)$ that can scatter to $(-\pi+\delta,0)$ due to partial nesting and the flat dispersion. To compensate for the momentum transfer, there are only few states on the Fermi surface. Those lie between $\theta=3/8\,\pi$ and $1/4\,\pi$ and have therefore extremely high scattering rates.}
\label{fig: angular distribution of the current}

\end{center}
\end{figure*}

We can also analyze the current distribution as a function of the temperature. In Fig. \ref{fig: angular distribution of the current} we display the angular distribution of the charge current for two different doping levels. At the lowest temperatures the distribution function changes its angular dependence strongly and the weight of the current is shifted from small angles to larger angles, while above a certain temperature the angular distribution remains invariant. 

These regimes can be associated with two different kinds of scattering mechanisms. While at low temperature impurity scattering dominates, for sufficiently high temperature the transport is mainly determined by particle-particle scattering. Since only umklapp scattering contributes noticeably to the resistivity, the two-particle scattering contribution is anisotropic, while the $\delta$-potential impurities give isotropic scattering. This difference accounts for a qualitative change in the angular dependence of the distribution function. We will return to the interplay of the two scattering mechanisms in the context of a discussion of Matthiessen's rule, Sec. \ref{sec: Momentum relaxation mechanism in the cuprates and Matthiessens rule}.

\subsection{Effect of the vertex renormalization}
\label{subsec: discussion of the parameters}

We turn now to the role of the renormalization of the scattering vertex in the unconventional nature of the transport properties.

As explained in App. \ref{sec: choice of parameters} we used a value of $U_{\Lambda}=2\,t$ for \lsco. When we extend our calculation to a series of lower initial values of $ U_{\Lambda} $ the temperature dependence of the resistivity and the coefficients of the corresponding linear plus quadratic fit are shown Fig. \ref{fig: rhoVsTemperatureLSCO} and in Fig. \ref{fig: coefficientsLSCO}. In addition we show the resistivity for the unrenormalized interaction $U_{\infty} = 2.0\,t $ which corresponds to an FRG treatment with infinite cutoff. In this case the scattering rates are determined only by the scattering phase space which is specified by the Fermi function and the geometry of the Fermi surface. 

Comparing the resistivity curves for $U_{\Lambda}=2.0\,t$ and $U_{\infty}=2.0\,t$, we observe very similar overall trends as function of doping. The high-temperature resistivity grows upon reducing the hole-concentration and the $T$-linear part increases at the same time. It is obvious that the scattering phase space plays an important role in the change of the behavior of resistivity for different hole-concentrations. Overall the behavior is, however, less pronounced for the unrenormalized than renormalized vertex as is clear in Fig. \ref{fig: coefficientsLSCO}. The renormalization of the vertex causes a sizable increase of the resistivity as one can see in the comparison of the curves for $U_{\Lambda}> 2.0\,t$ and $U_{\infty}=2.0\,t$, and it also strongly enhances the trend toward linear resistivity. This feature is also easily identified from the $U$-dependence of the quadratic term in Fig. \ref{fig: coefficientsLSCO}. For stronger initial interaction, and therefore stronger renormalization effects, the quadratic term is substantially reduced, even as the total resistivity, the sum of linear and quadratic term is increased. Thus, it is evident that correlation effects included in the vertex renormalization by the FRG scheme, make a substantial contribution of the unconventional transport properties, in particular, in the regime close to optimal doping.

The vertex renormalization leads to a non-trivial momentum space structure of the effective scattering vertex which can be translated to a longer ranged real space interaction compared to the local Hubbard-$U$. As a second effect of the FRG analysis, the momentum space structure but also the absolute values of the scattering vertex become temperature dependent. In order to discriminate between these two effects we modified our calculations in a way as to include only the momentum space renormalization of the vertex, but neglect the temperature dependence. This was implemented by stopping the RG flow with a large cutoff $\Lambda$. In this case \emph{large} refers to the cutoff exceeding the largest temperature of our study. While this completely suppresses the temperature dependence of the scattering vertex, it still introduces momentum space anisotropy.

The result resembles the behavior of the simple unrenormalized onsite repulsion $U_\infty=2.0d\,t$ of Fig. \ref{fig: rhoVsTemperatureLSCO}. The introduction of momentum anisotropy through the RG flow has almost no qualitative effect on the temperature dependence of the resistivity. From this comparison we can conclude that the pronounced linear resistivity in the limit of $T\rightarrow 0$ is a consequence of the strongly temperature dependent quasiparticle scattering rates. This supports the idea of Ref. [\onlinecite{ossadnik2008}] where the unconventional temperature dependence of the scattering rates through the RG flow were described. 

At high temperature, however, the growth of the linear-temperature term when the doping is decreased toward optimal doping, is mainly due to the Fermi surface geometry and the temperature dependence of the scattering volume, while the momentum anisotropy of the scattering interaction is responsible for the enhanced value of the resistivity. 

\section{Momentum relaxation mechanism and Matthiessen's rule}
\label{sec: Momentum relaxation mechanism in the cuprates and Matthiessens rule}

In the cuprates the resistivity is caused by electron-electron interactions which raises the question, how the total momentum of the entire distribution function is relaxed. In the following we will discuss which contributions are combined in the resistivity calculations of the previous section.

\subsection{Impurity and umklapp scattering}
\label{subsec: Umklapp scattering}

In impurity scattering processes, the momentum difference between initial and final state is transferred to the lattice and determines the residual resistivity as the zero-temperature limit. A deeper analysis shows that the resistivity from impurity scattering can have a weak temperature dependence due to thermal broadening of the distribution function and a possible bias toward higher or lower quasiparticle velocity. In our calculation the impurities are modeled as $\delta$-potentials. 

A second and more complex contribution to the electrical resistivity originates from electron-electron scattering with finite momentum transfer 
to the lattice, so-called umklapp scattering, where a momentum corresponding to a reciprocal lattice vector is transferred to the lattice. Note that the Fermi surface has to be sufficiently large in order to allow for umklapp assisted scattering between different momenta on the Fermi surface.

For a square lattice we can distinguish between two kinds of umklapp scatterings: with momentum transfers of the type $(2\pi/a,0)$ and $(2\pi/a,2\pi/a)$ to the lattice. The first kind of umklapp scattering has a large phase space for a sufficiently large Fermi surface with regions where the $x$-component of the Fermi vector satisfies the condition $ |k_{Fx}| \leq \pi/2a$ and other regions with $|k_{Fx}|\geq \pi/2a$. The second type of umklapp process is only possible, if the Fermi surface intersects with the so-called umklapp surface, which is the rotated square (diamond) that connects the four saddle points at $(0,\pm\pi/a)$ and $(\pm\pi/a,0)$. This umklapp scattering has relatively small phase space due to the stringent momentum constraints.
Within our model for \lsco\ the $(2\pi/a,0)$ umklapp process largely dominates over the $(2\pi/a,2\pi/a)$ umklapp process for all doping levels as far as the charge transport resistance is concerned. On the strong overdoped side, $(2\pi/a,2\pi/a)$ umklapp scattering is even completely absent in the low-temperature limit.

Due to the geometrical constraints for umklapp scattering the momentum space structure of the scattering vertex can play an important role. The temperature dependence of the resistivity is controlled by umklapp scattering. If the scattering rates become especially high on certain parts of the Fermi surface, their contributions to transport will be suppressed yielding the possibility for unconventional temperature scaling.

\subsection{Deviations from Matthiessen's rule}

Often different contributions to the resistivity are assumed to be decoupled and one adds them as serial resistors. This decoupling leads to Matthiessen's rule, which  is only justified, if each individual contribution can be well described by a single relaxation time. In this case the collision integral in the Boltzmann equation directly splits into the scattering channels. In the general case, when such a single relaxation time approximation is insufficient, the collision integral does not decouple.

In Fig. \ref{fig: angular distribution of the current} we have seen that in the low-temperature regime the angular distribution of the current changes its form. This was attributed to a crossover from an impurity scattering dominated regime to a two-particle scattering regime. These different scattering mechanisms lead to different current distributions due to their difference in the scattering geometry. While impurity scattering is isotropic, the umklapp processes responsible for the momentum relaxation in the two-particle scattering channel are highly anisotropic leading to a non-trivial interplay of impurity scattering with two-particle interactions and a potential breakdown of Matthiessen's rule.

To examine the deviations from Matthiessen's rule we compare our calculation with the sum of the ''individual'' contributions to the resistivity. First we remove the impurities to obtain a ''bare'' electron-electron resistivity $ \rho_{\rm ee}$, second we estimate $\rho_{\text{imp}}$ by considering free (non-interacting) electrons scattered by impurities. 

In a first step we compare the linear and the quadratic coefficients of a $T+T^2$-fit for the clean limit, $\rho_{ee}$, to the coefficients in the original calculation including impurities of Fig. \ref{fig: coefficientsLSCO}. The quadratic coefficient grows weakly down to $p\approx 0.25$ and remains constant at $p<0.25$, while for the full calculation the quadratic coefficient decreased below $p\approx0.30$ and even went to zero for sufficiently strong coupling. Furthermore, the linear term is somewhat smaller without impurities. This difference in the behavior of the coefficients is a violation of Matthiessen's rule.

\begin{figure} [t]
\begin{center}

\includegraphics[width=\columnwidth]{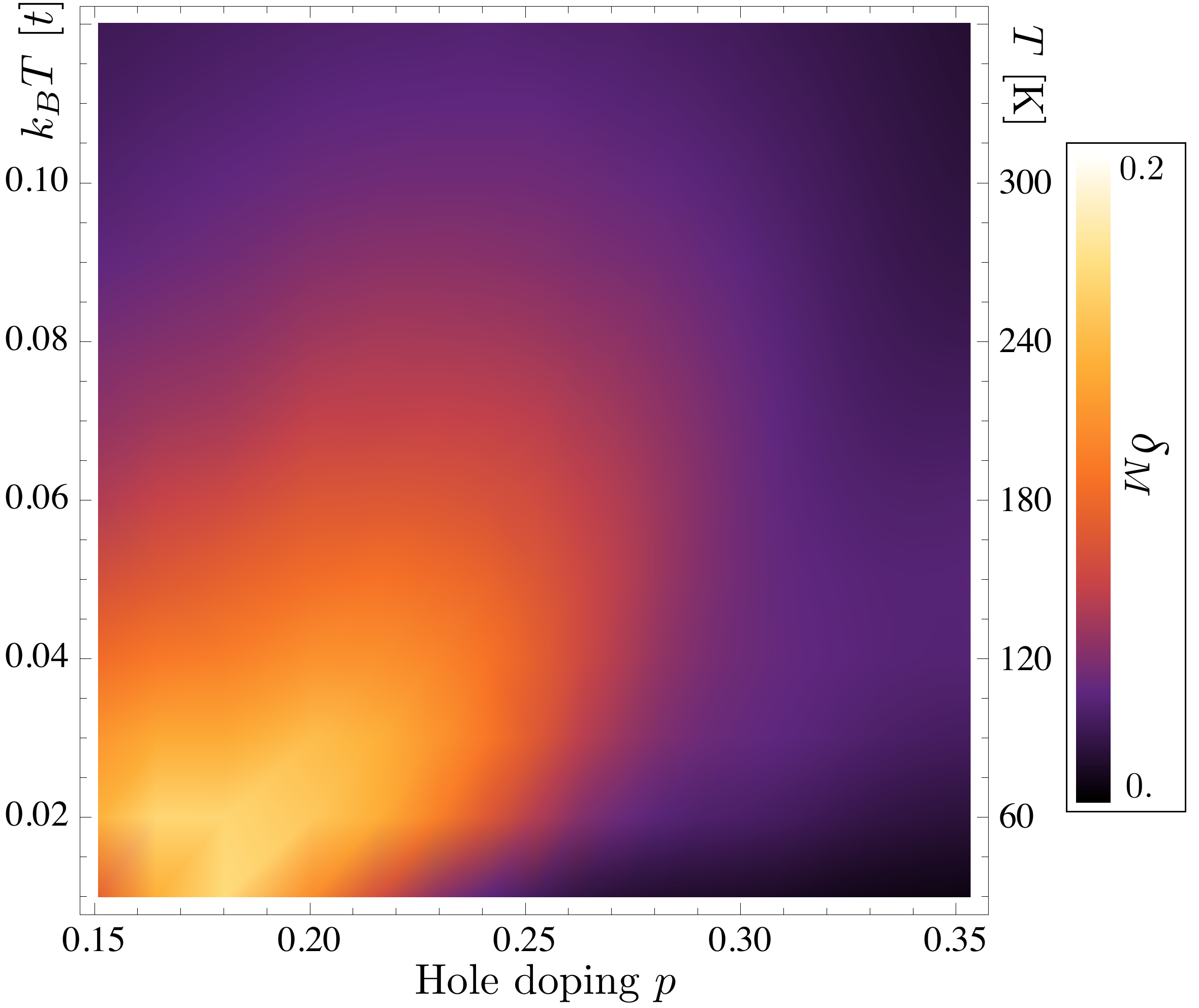}
\caption{Deviation from Matthiessen's rule: We subtract the sum of the resistivity for pure impurity scattering and pure electron-electron scattering from the resistivity of the full calculation $\rho_{\text{full}}$ and normalize it to $\rho_{\text{full}}$. It is clearly visible that the deviation $\delta_{M}$ (cf. Eq. \eqref{eq: deviation from Matthiessens rule}) is most pronounced in the critical region of the phase diagram.}
\label{fig: matthiessenRuleDeviation}

\end{center}
\end{figure}

In order to make this more evident, we plot the deviation from Matthiessen's rule, $\delta_{M}$, quantitatively defined as 
\begin{align}
	\delta_{M}=[\rho_{\text{full}}-(\rho_{\text{ee}}+\rho_{\text{imp}})]/\rho_{\text{full}},
\label{eq: deviation from Matthiessens rule}
\end{align}
as a function of the temperature for different doping levels. The result is shown in a color plot in Fig. \ref{fig: matthiessenRuleDeviation}. We immediately observe that the deviation from Matthiessen's rule is strongest at the optimally doped regime at low temperature. The difference in the scattering geometry for impurity scattering and two-particle scattering is most pronounced in this part of the diagram. Here the deviation $\delta_{M}$ reaches values as large as 16\%.

It should be noted that the one-loop FRG approximation is less reliable at low temperature and low doping, as pointed out earlier. Hence the magnitude of $\delta_{M}$ in our calculation carries considerable uncertainties in this part of the diagram. Nevertheless, the violation of Matthiessen's rule is an generic property for systems featuring a crossover from isotropic impurity scattering to strongly anisotropic inter-particle scattering dominated transport.

\section{The Seebeck effect in overdoped \lsco}
\label{sec: the seebeck effect in overdoped lsco}

The Seebeck effect is a further example which allows us to test Fermi liquid properties in transport. This effect as the electric response to a temperature gradient is based on thermo-diffusion of charge carriers and can be analyzed with Boltzmann transport theory too. 

\subsection{Experimental observation of the Seebeck effect in \textrm{Eu}-\lsco}

\begin{figure*}[t]
\begin{center}

\includegraphics[width=0.45\textwidth]{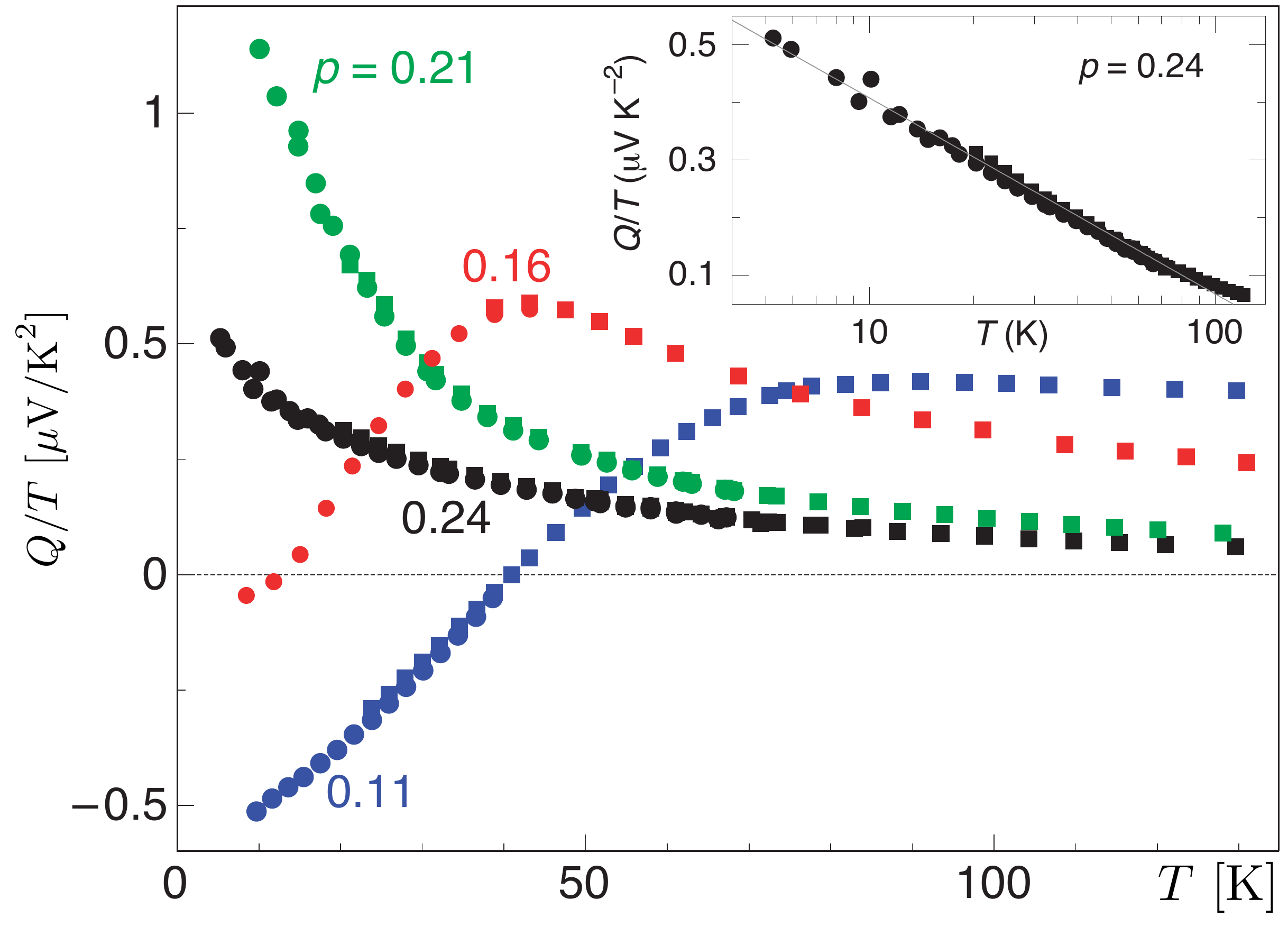}\qquad
\includegraphics[width=0.47\textwidth]{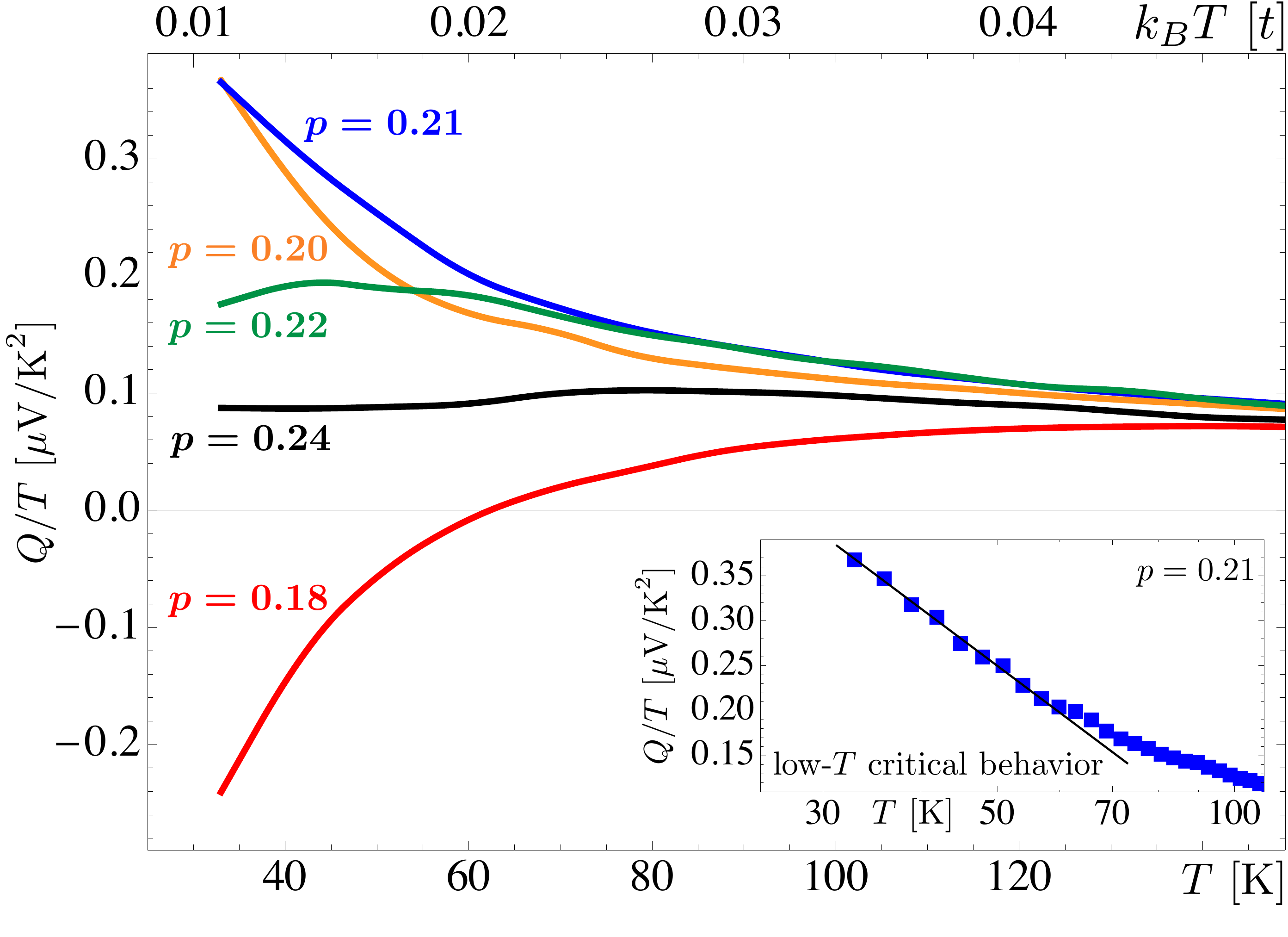}
\caption{\emph{left}: Experimental data of the Seebeck coefficient $Q$ normalized to the temperature for Eu-LSCO ($\textrm{La}_{1.8-x}\textrm{Eu}_{0.2}\textrm{Sr}_{x}\textrm{CuO}_{4}$), adapted from Ref. [\onlinecite{laliberte2011}]. \emph{right}: Numerical calculation of the Seebeck coefficient $Q$ normalized to the temperature for our model of \lsco. Note that the temperature range does not extend to zero.}
\label{fig: experimental seebeck eu-lsco and numerical seebeck}

\end{center}
\end{figure*}

Lalibert\'{e} et al. measured the Seebeck coefficient $Q$ of $\textrm{La}_{1.8-x}\textrm{Eu}_{0.2}\textrm{Sr}_{x}\textrm{CuO}_{4}$ [Eu-\lsco], cf. Ref. [\onlinecite{laliberte2011}]. While their study aims at probing the Fermi surface reconstruction scenarios in the underdoped regime of the phase diagram, they also present data up to a hole-doping level of $p=0.24$. Data of the Seebeck coefficient are displayed in Fig. \ref{fig: experimental seebeck eu-lsco and numerical seebeck} (left panel) for the doping regime $p>0.11$.  Europium doping of \lsco\ is believed to have a stronger influence on the properties of underdoped samples, but little effect is expected in the overdoped regime. 

For underdoped samples $Q/T$ shows a sign change, starting negative at low temperature it turns positive at higher temperature. On the other hand, for overdoped samples $Q/T$ remains positive for all temperatures measured and shows
a rather strong increase toward low temperature near optimal doping. The inset of Fig. \ref{fig: experimental seebeck eu-lsco and numerical seebeck} (left panel) demonstrates the $\log(1/T)$ dependence  of $ Q/T $ at low temperature, which has been interpreted
as a signature of quantum criticality.

\subsection{Numerical results}

\begin{figure*}[t]
\begin{center}

\includegraphics[width=0.49\textwidth]{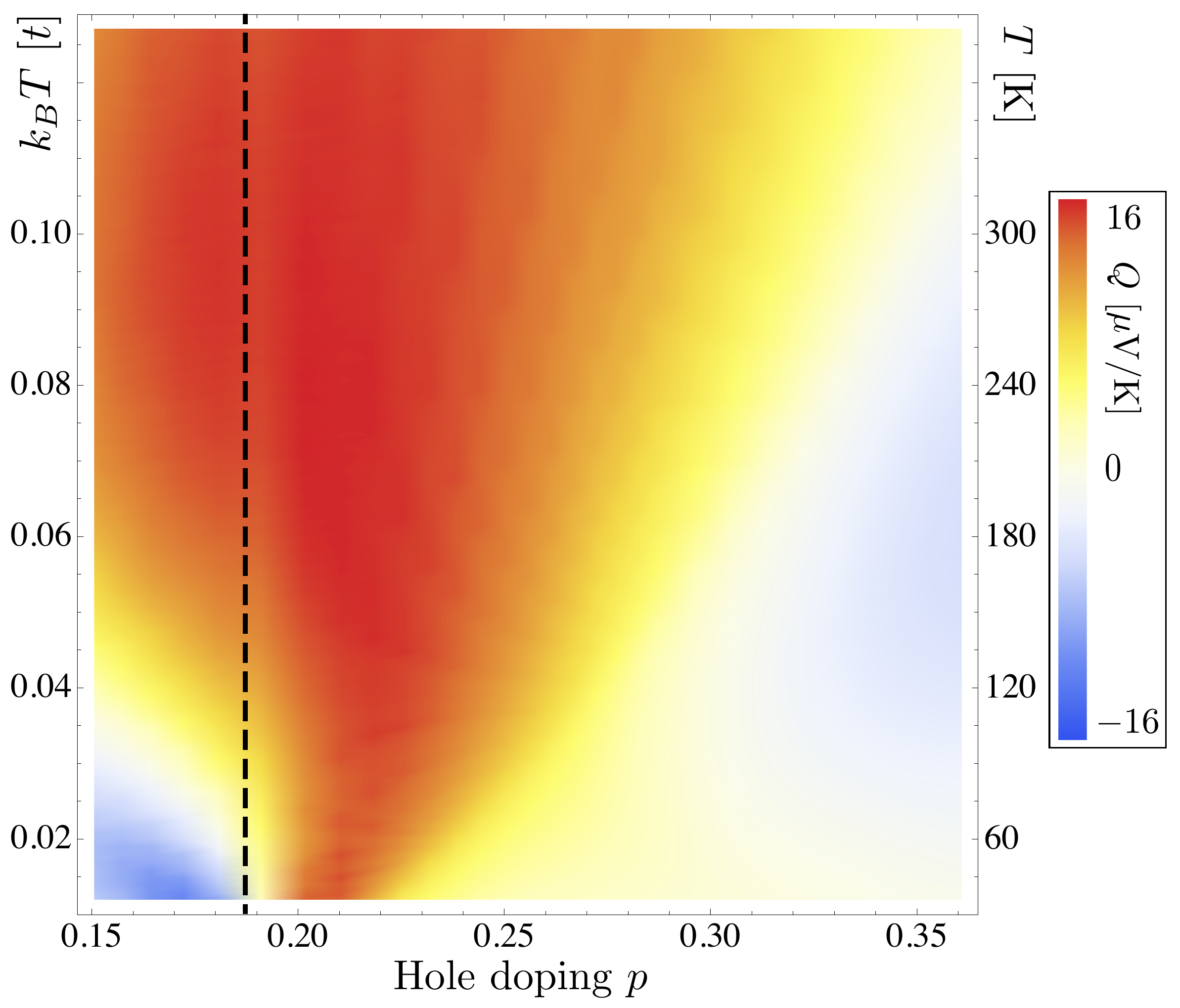}
\includegraphics[width=0.49\textwidth]{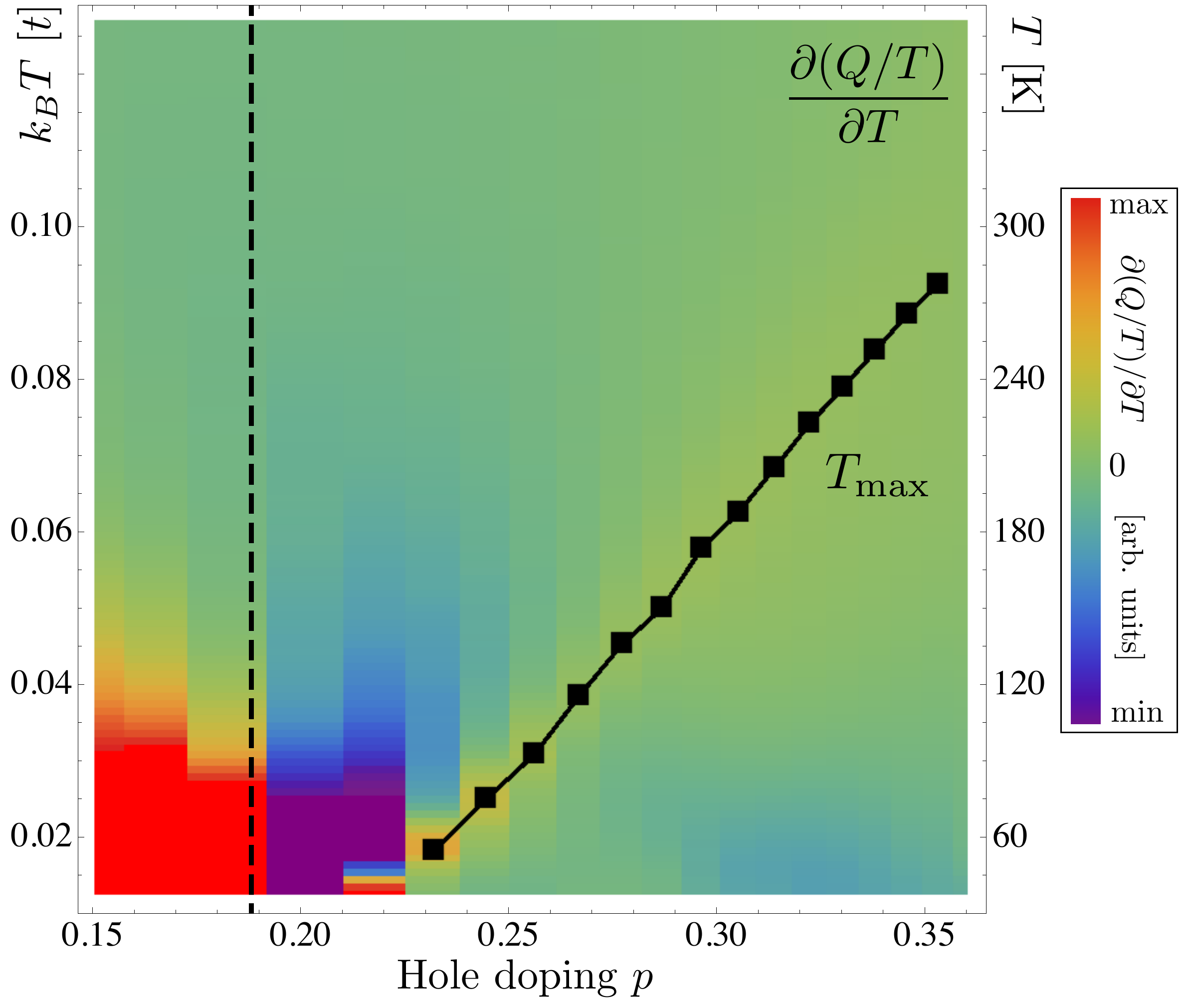}
\caption{\emph{left}: Seebeck coefficient as a function of hole doping and temperature. \emph{right}: Derivative of $Q/T$ with respect to the temperature: Fermi liquid theory predicts a linear temperature dependence of the Seebeck coefficient. Thus, any non-zero value of this derivative indicates deviations from Fermi liquid theory.}
\label{fig: seebeck phase diagram}

\end{center}
\end{figure*}

We have numerically calculated the Seebeck coefficient for \lsco\ within our semiclassical approach, as explained in App. \ref{subsec: calculation of the seebeck coefficient}. The results are shown in Figs. \ref{fig: experimental seebeck eu-lsco and numerical seebeck} (right panel) and \ref{fig: seebeck phase diagram}. A good overview on the behavior of the Seebeck coefficient as 
a function of doping and temperature is obtained from Fig. \ref{fig: seebeck phase diagram} (left panel). At the doping $ p_c  \approx 0.19 $ we observe a sign change as we cross the Lifshitz transition at low temperature. Interestingly, for $p<p_{c}$ increasing the temperature leads to a sign change from negative at low to positive at higher temperature. Turning to the other limit of strong overdoping the Seebeck coefficient turns negative again as expected for electrons is a band with small filling. We also observe a doping dependent temperature scale where $Q$ displays a rapid increase. Interestingly, this temperature compares well with $T_{1}$.

In order to compare our numerical with the experimental results, we have plotted $ Q/T $ in Fig. \ref{fig: experimental seebeck eu-lsco and numerical seebeck} for some selected values of hole doping. The inset shows the low-temperature regime for the doping level $p=0.21$ on a logarithmic temperature scale.

The general structure of the calculated Seebeck coefficient is in good qualitative agreement with the experimental data measured for Eu-\lsco\ Fig. \ref{fig: experimental seebeck eu-lsco and numerical seebeck}. In both cases, there is a critical doping level below which the Seebeck coefficient changes sign with increasing temperature, starting from negative in the $T\rightarrow 0$ limit. Above this critical doping level, $Q/T$ becomes very large for low temperature and strongly decreases with temperature. The low-temperature rise of $Q/T$ becomes less pronounced with increasing hole doping. In the low-temperature regime, $Q/T$ acquires a $\log(1/T)$ dependence which is clearly visible in both the experimental data and the numerical data below $60\,\textrm{K}$. 

A small discrepancy between our model and the experiment is the value of the critical doping level. We have designed the band structure that we use to describe \lsco\ such that the Lifschitz transition is located at $p_{c}\approx 0.19$, cf. App. \ref{sec: model for the quasiparticle dispersion of lsco}. The van Hove singularity at the Lifschitz transition is the origin for the critical behavior in our model and thus constrains the critical doping level to $p_{c}$. In the experimental study, however, a critical doping level of $0.24$ is reported.

\subsection{Signature of critical transport properties}

In the last section, we have discussed the low-temperature $\log(1/T)$ dependence of $Q/T$ and the sign changes with doping or temperature. These properties of the Seebeck coefficient are often taken as indication for the presence of a quantum critical point [\onlinecite{QCPfromSeebeckMeasurements,custers2003}]. These features are found in both the experimental data and the numerical simulation. We now want to locate the areas of critical behavior in the phase diagram. 

For conventional metals, the Seebeck coefficient is a linear function of the temperature manifesting Fermi liquid behavior. 
Deviations from this linear dependence may be used as a signature for non-Fermi liquid physics. Thus, we analyze the the temperature derivative of $ Q/T $ which should vanish for Fermi liquids. 
Fig. \ref{fig: seebeck phase diagram} (right panel) displays this function. This value is largest at low temperature around optimal doping and a sign change appears at the Lifshitz transition. This is the region where the strongest differences with the Fermi liquid picture emerge. From this finding we may identify the vicinity of the van Hove singularity and the presence of a Lifshitz transition as the origin of non-Fermi liquid behavior and the signatures of quantum criticality.  

Interestingly, Fig. \ref{fig: seebeck phase diagram} (right panel) allows us also to see the temperature of the sharp increase of
$Q/T$, cf. Fig. \ref{fig: seebeck phase diagram} (left panel), appearing as a maximum of $\partial (Q/T) / \partial T$. This characteristic temperature $T_{\text{max}}$ 
separates the high-doping low-temperature regime coming closest to Fermi liquid behavior ($ T^{2}$-dependence of resistivity)
from the lower-doping high-temperature regime which is characterized by the non-Fermi liquid $T$-linear resistivity, as discussed in Sec. \ref{subsubsec: scaling regimes}.

\section{Discussion and Conclusion}

We investigate the temperature and doping dependence of charge transport coefficients using a combination of an experimentally determined band structure and a renormalized scattering matrix calculated in a one-loop renormalization group approximation. The band dispersion matches the quasiparticle spectrum  in ARPES for LSCO [\onlinecite{yoshida2006}], and as such includes the renormalization of the dispersion with hole density due to interactions. The electron-electron scattering vertex develops momentum-space and temperature dependence as the hole density is reduced. We have chosen a moderate to weak value for the bare onsite Coulomb repulsion consistent with the one-loop approximation. We also refrain from including the pseudogap region which would require a solution of the strong coupling regime that results from divergences in the FRG flow.
The Boltzmann transport equation is solved numerically with collision integrals for electron-electron and impurity scattering determined by Fermi's Golden rule. This semiclassical approach determines the quasiparticle distribution functions taking into account the full angular and energy dependence of the collision integral.

We find  good qualitative agreement with the main trends in the experiments on the in-plane resistivity of \lsco\  [\onlinecite{hussey2009}]. This includes the increase of the temperature dependent term in the resistivity upon decreasing the hole-concentration and the trend toward a linear temperature dependence of the resistivity $ \rho(T) $ around optimal doping. A parameterization of our results using a polynomial fit in $T$ up to second order, inspired by the experimental analyses [\onlinecite{hussey2009,greven2012}], gives  consistency between the trends in experiment and our model. While the $T$-linear behavior of $ \rho(T) $ is often attributed to the existence of a quantum critical point around optimal doping [\onlinecite{mackenzie2001,dagan2004}], our discussion shows that this feature may also be understood as a consequence of the unusual temperature dependence of the renormalized scattering vertices and the increased scattering phase space around this doping regime, where a Lifshitz transition of the band structure leads the chemical potential very close to a van Hove singularity. The magnitude of the calculated resistivity is substantially lower than in the experiment, consistent with the moderate to weak interaction values and one-loop RG approximation

Motivated by the experimental data analysis [\onlinecite{hussey2011}] we also identified crossover regimes in  $ d \rho(T)/dT $ and find rather
sharp changes in the temperature dependence. There are two characteristic temperatures $ T_{1} $ and $T_{2}$ where the latter denotes the lower bound of the $ T$-linear regime of $ \rho(T) $ and $ T_{1} ( < T_{2}) $ the upper bound for a purely $ T^{2} $-dependence (Fermi liquid regime). The intermediate temperature range may be considered as combination of both. The resulting phase diagram gives a 
very good account of the trends  of $ \rho(T) $ observed in the experiment. In particular, we can show that there is a clear correlation of the unconventional temperature scaling with
the van Hove singularity, quantified through the effective distance of the chemical potential from the van Hove level, $ k_B T_{\Delta vH} = 4 | \mu - E_{vH}| $, cf. Fig. \ref{fig: ExponentPhaseDiagram}.

By comparing the temperature dependence of $\rho(T)$ for renormalized and bare scattering rates, we could demonstrate the importance of the scattering vertex renormalization. While in the strongly overdoped regime, the difference in the numerical data determined from renormalized and unrenormalized scattering rates is not very pronounced, around optimal doping the phase space driven growth of a linear term is strongly enhanced due to renormalization of the quasiparticle interactions. 

It also turns out that the impurity scattering is important in this respect as well, contributing to the unconventional temperature dependence. This is surprising as impurities contribute usually only a weakly  temperature dependent offset to resistivity. We found deviations from Matthiessen's rule in large parts of the considered phase diagram. This deviation is especially pronounced as we approach optimal doping at intermediate to low temperature due to the different momentum dependence of impurity and electron-electron scattering. Matthiessen's rule, however, is recovered in the more conventional, strongly overdoped regime at low temperature.

We have also studied the Seebeck coefficient within our model. We find qualitative agreement between our numerical data and the main trends in the measurements performed on a slightly different compound, Eu-\lsco. We believe that  the Eu-doping does not strongly influence the Seebeck coefficient in the optimal to overdoped doping regime which we are interested in. The calculated Seebeck coefficient, also displays a critical region in the phase diagram with pronounced deviations from Fermi liquid behavior. We have related the critical behavior to proximity of the van Hove singularity to the Fermi energy which is linked to the Lifschitz transition.

\begin{acknowledgements}

We are grateful to Carsten Honerkamp, Nigel Hussey, L\'{a}szl\'{o} Forr\'{o}, and Stevan Arsenijevi\'{c}  for helpful discussions. This study was supported by the Swiss Nationalfonds, the NCCR MaNEP, the HITTEC project of the Competence Center Energy \& Mobility and the Synergia TEO. 

\end{acknowledgements}

\begin{appendix}

\section{The quasiparticle dispersion of \lsco}
\label{sec: model for the quasiparticle dispersion of lsco}

We use a band structure parametrized within a tight-binding model on a square lattice, including nearest, next-nearest, and next-to-next-nearest neighbor hopping
\begin{align}
	\varepsilon_{\bvec{k}} & = -2t(\cos k_{x}+ \cos k_{y}) + 4t' \cos k_{x}\cos k_{y} \nonumber \\
	&\phantom{=}\quad - 2 t'' (\cos 2k_{x} + \cos 2k_{y}).
\label{eq: tight binding model}
\end{align}
The hopping parameters $t$, $t'$, $t''$, and the chemical potential $\mu$ define the geometry of the Fermi surface and the filling of the band.

Based on their ARPES study, Yoshida et al. [\onlinecite{yoshida2006}] have fitted the low-energy quasiparticle spectrum of \lsco\ and the Fermi surface in the doping range from $p=0.03$ to $p=0.30$ using the model \eqref{eq: tight binding model}. Fixing 
he magnitude of $t$ to $0.25\,\textrm{eV}$ and the ratio $t''/t'=0.5$ for all dopings they used $\mu$ and $t'$ as fitting parameters.  Their results are shown in Fig. \ref{fig: discretization}. The relation between $\mu$ and $t'$ as a function of the doping is linear to a very good approximation,

\begin{align}
	\mu(p) = -1.85t + 7.26\,t'(p). 
\label{eq: renormalization scheme}
\end{align}
Yoshida et al. identified a Lifshitz transition at $p\approx 0.18$. Our hopping parameter renormalization scheme, Eq. \eqref{eq: renormalization scheme}, locates the Lifshitz transition at $p\approx 0.24$. Changing $t''/t'$ to $t''/t'=0.25$ shifts the Lifshitz transition within our model to the experimental doping level. Note, that a second ARPES study by Ino et al. [\onlinecite{ino2002}] also confirms the presence of a Lifschitz transition at a doping level of about $p\approx 0.20$.

\section{Strongly renormalized quasiparticle interactions}
\label{sec: Strongly renormalized quasiparticle interactions}

\subsection{Renormalization of the scattering vertex} 

\begin{figure*}[t]
\begin{center}

\includegraphics[width=1\textwidth]{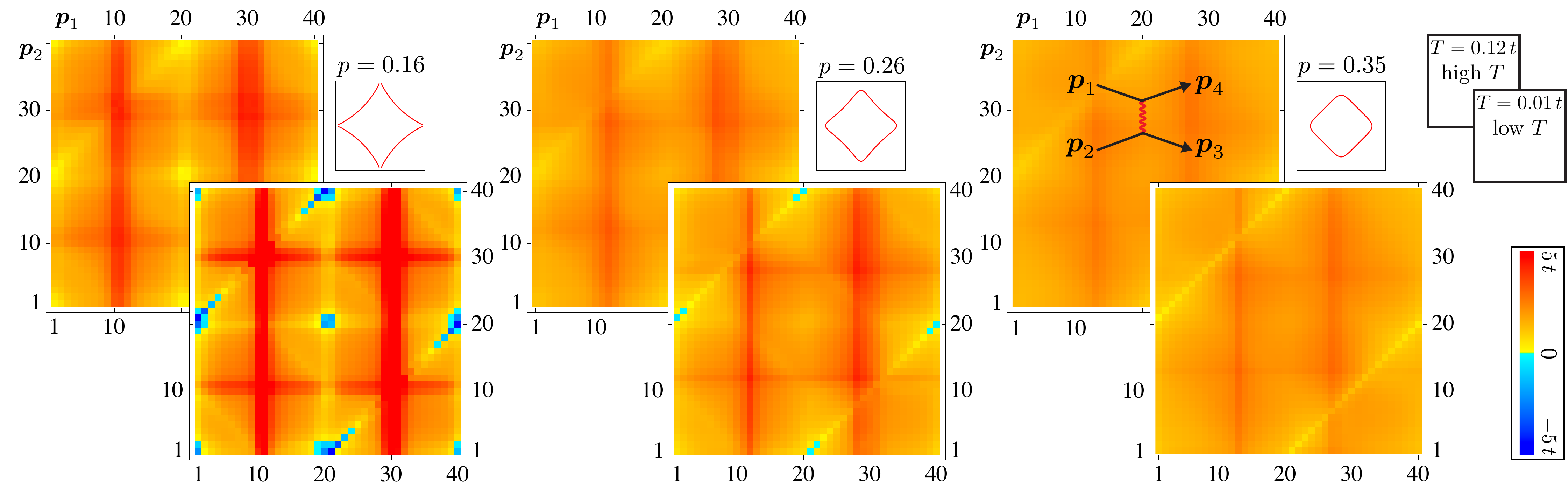}
\caption{The renormalized vertex $V(\bvec{p}_1,\bvec{p}_2,\bvec{p}_3,\bvec{p}_4)$ as a function of the momenta $\bvec{p}_{1}$ (horizontal-axis) and $\bvec{p}_{2}$ (vertical-axis) for fixed $\bvec{p}_{3}$ in the first angular patch. All three momenta lie on the Fermi surface and are therefore uniquely determined by their angular patch label. The fourth momentum variable $\bvec{p}_{4}$ is fixed by momentum conservation and does in general not lie close to the Fermi surface. We demonstrate the doping and temperature dependence by showing the vertex for three different values of the hole-doping $p$ for low respectively high temperature. The initial value of the onsite repulsion is $U=2.0\,t$ and the RG-cutoff scale $\Lambda_{c}=0.01\,t$. The angular discretization scheme (here 40 patches) is illustrated in Fig. \ref{fig: discretization} (a).}
\label{fig: vertex}

\end{center}
\end{figure*}

The calculation of the renormalized vertex in Ref. [\onlinecite{ossadnik2008}] is based on the FRG equation for the one-particle irreducible generating functional derived in Refs. [\onlinecite{honerkamp2001a,honerkamp2001b}] which gives hierarchically coupled flow equations for the one-particle irreducible vertices. The RG-flow follows from a Wilsonian flow scheme with a sharp momentum cutoff. The RG equations are solved using the standard truncation of all vertices with more than four legs, so that the self energy and the four-point vertex are the only remaining quantities in the calculation. For the vertex, the momentum dependence is discretized and constrained to momenta lying on the Fermi surface. The frequency dependence of the vertex is suppressed. For the RG-flow, the flow parameter $\Lambda$ is taken from $\infty$ to zero. In general, the truncated flow tends to diverge for low temperature at finite energy scales $\Lambda$, indicating the appearance of an instability. The energy scale at which the divergence occurs is related to the critical temperature $T_{c}$ of the corresponding instability, in our case superconductivity. Note, however, that due to the truncation of the RG equations the $T_{c}$ is largely overestimated. 

For a study of normal state transport, the divergence of the vertex should be avoided. While in the experiment superconductivity is suppressed by a magnetic field which is difficult to incorporate in the RG analysis, we keep thge cutoff energy scale $\Lambda_{c}>0$ finite and so suppress the leading $d$-wave pairing instability in the RG flow. The choice for $\Lambda_{c}$ is discussed in App. \ref{sec: choice of parameters}.

The quasiparticle scattering vertex $V(\bvec{p}_1,\bvec{p}_2,\bvec{p}_3,\bvec{p}_4)=\nbra{\bvec{p}_{1}\bvec{p}_{2}}\hat{V}\nket{\bvec{p}_{3}\bvec{p}_{4}}$ is peaked for momenta $\bvec{p}_{i}$ lying close to the saddle-points $(0,\pi)$ and $(\pi,0)$ and has especially strong contributions for momentum transfer of $(\pi,\pi)$, cf. Fig. \ref{fig: vertex}. Further discussions of the RG flow and interpretations of vertex diagrams as in Fig. \ref{fig: vertex} are given in Refs. [\onlinecite{honerkamp2001a,honerkamp2001b,honerkamp2001c,honerkamp2001d}].

The FRG method is a weak coupling analysis. On the underdoped side, below a temperature scale $T^{*}$ a pseudogap opens as an indication that the system is driven into a strong coupling phase. The strong-correlation physics are not included in our analysis. Hence, we restrict our study to dopings from optimal to strongly overdoped.

\subsection{Choice of parameters}
\label{sec: choice of parameters}

The onsite repulsion $U$ for the cuprates is a large energy scale of the order of the band width as at half-filling a Mott insulating phase is realized. Using such large values of $U$, however, leads to a divergence of the RG flow within the normal state temperature range, since the instability temperatures are overestimated. This urges us to start with moderate value of the onsite repulsion of $U\approx 2.0\,t$ which then is renormalized to much larger values for the low-energy quasiparticles after integrating out the high-energy states. Moreover, we suppress the $d$-wave pairing instability by keeping the cutoff $\Lambda_{c}>0$ finite in the RG flow, as mentioned above. For this purpose we choose $\Lambda_{c}=k_{B}T_{\text{min}}$ corresponding to the lowest temperature in our study.

\section{Semiclassical treatment of normal state charge transport of \lsco}
\label{sec: Effective Model for the Normal State Charge Transport of the High-Temperature Superconductor LSCO}

\subsection {The Boltzmann equation and its discretized solution}

\begin{figure*}[t]

\includegraphics[width=\textwidth]{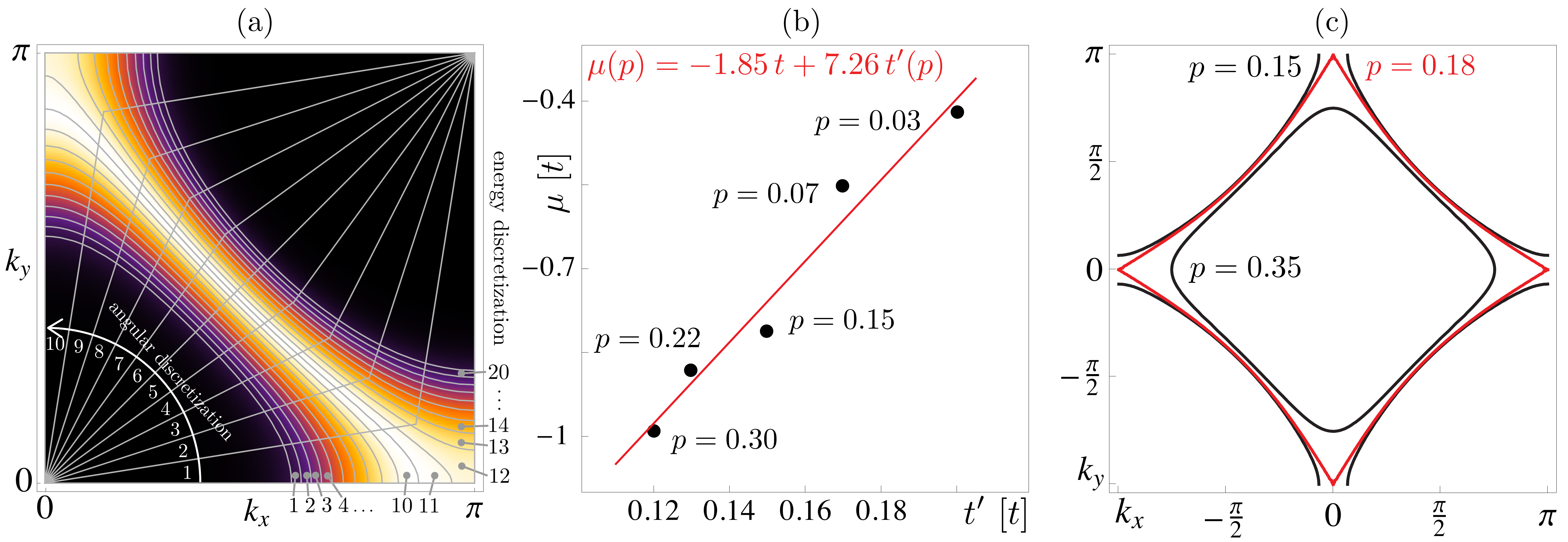}

\caption{(a) Sketch of our curved-patch discretization of first quarter of the Brillouin zone. Plotted in color-code is the $f_0(\bvec{p})(1-f_0(\bvec{p}))$, a measure of the scattering phase space. The lines represent the edges of the finite elements (boxes). For illustration purposes the resolution of the mesh is much lower than used for our calculations.
(b) The points represent the relation between $\mu$ and the next-nearest neighbor hopping $t'$ observed in the experiment [\onlinecite{yoshida2006}]. We use the linear fit (red line) to this data as the renormalization scheme for the hopping parameters in our model.
(c) Fermi surface of the tight binding model \eqref{eq: tight binding model} with hopping parameters chosen to match the measured parameters for \lsco\ for different hole-doping values, cf. App. \ref{sec: model for the quasiparticle dispersion of lsco}. 
}
\label{fig: discretization}

\end{figure*}

The Boltzmann equation relates the substantial derivative of the quasiparticle distribution function to the collision integral, 
taking all scattering events into account which yield a change of the distribution function,
\begin{align}
	\frac{\text{d}f(\bvec{r},\bvec{p},t)}{\text{d}t}&=\frac{\partial f(\bvec{r},\bvec{p},t)}{\partial t}+\bvec{v}(\bvec{p})\cdot\bvec{\nabla}_{r}f(\bvec{r},\bvec{p},t)\nonumber\\
	&\phantom{=}\quad-\left[\bvec{\nabla}_{r}V(\bvec{r})\right]\cdot\bvec{\nabla}_{p}f(\bvec{r},\bvec{p},t)\phantom{\int}\nonumber\\
	&= \left.\frac{\partial f(\bvec{r},\bvec{p},t)}{\partial t}\right|_{\text{coll}}.
\label{eq: The Boltzmann Equation}
\end{align}
We consider two contributions in the collision integral, two-particle 
collisions labeled by "ee" and impurity scattering labeled by "imp":
\begin{align}
	\left.\frac{\partial f(\bvec{p})}{\partial t}\right|_{\text{coll}} = \left.\frac{\partial f(\bvec{p})}{\partial t}\right|_{\text{ee}} + \left.\frac{\partial f(\bvec{p})}{\partial t}\right|_{\text{imp}}
\label{eq: separation of the collision integral into EE and IMP}
\end{align}

The first contribution is expressed as, 
\begin{align}
	\left.\frac{\partial f(\bvec{p}_{1})}{\partial t}\right|_{\text{ee}}&= -\int \text{d}\bvec{p}_{2}\,\text{d}\bvec{p}_{3}\,\text{d}\bvec{p}_{4}\;\Gamma^{\text{ee}}_{1,2,3,4}\;F_{1,2,3,4},
\end{align}with\begin{align}
	\Gamma^{\text{ee}}_{1,2,3,4} &= \frac{1}{2}\left|\nbra{\bvec{p}_{1}\bvec{p}_{2}}\hat{V}\nket{\bvec{p}_{3}\bvec{p}_{4}}-\nbra{\bvec{p}_{1}\bvec{p}_{2}}\hat{V}\nket{\bvec{p}_{4}\bvec{p}_{3}}\right|^{2}\nonumber\\
	&\phantom{=}\times\delta(\bvec{p}_{1}+\bvec{p}_{2}-\bvec{p}_{3}-\bvec{p}_{4})\nonumber\\
	&\phantom{=} \times \frac{2\pi}{\hbar}\,\delta(\varepsilon(\bvec{p}_{1})+\varepsilon(\bvec{p}_{2})-\varepsilon(\bvec{p}_{3})-\varepsilon(\bvec{p}_{4})),
\intertext{and}
	F_{1,2,3,4}&=f(\bvec{p}_{1})f(\bvec{p}_{2})(1-f(\bvec{p}_{3}))(1-f(\bvec{p}_{4}))\nonumber\\
	&\phantom{=}-(1-f(\bvec{p}_{1}))(1-f(\bvec{p}_{2}))f(\bvec{p}_{3})f(\bvec{p}_{4})
\label{eq: collision integral EE}
\end{align}
where the transition rates $\Gamma$ are generated through Fermi's golden rule from the matrix elements of the quasiparticle interaction vertex $\hat{V}$, cf. App. \ref{sec: Strongly renormalized quasiparticle interactions}.
The second contribution to the collision integral is simpler due to its single particle form,
\begin{align}
	\left.\frac{\partial f(\bvec{p}_{1})}{\partial t}\right|_{\text{imp}}&= -\int \text{d}\bvec{p}_{2}\,\Gamma^{\text{imp}}_{1,2}F_{1,2},
\end{align}with\begin{align}
	\Gamma^{\text{imp}}_{1,2} &= n_{\text{imp}}\;|\! \bra{\bvec{p}_{1}}\hat{W}_{\text{imp}}\ket{\bvec{p}_{2}}\! |^{2}\;\frac{2\pi}{\hbar}\delta\left(\varepsilon(\bvec{p}_{1})-\varepsilon(\bvec{p}_{2})\right)
\intertext{and}
	F_{1,2}&=f(\bvec{p}_{1})(1-f(\bvec{p}_{2}))-(1-f(\bvec{p}_{1}))f(\bvec{p}_{2}).
\label{eq: collision integral IMP}
\end{align}

We ignore electron-phonon contributions here because, in the cuprates, the momentum relaxation of the quasiparticles is dominated by electron-electron interactions as was concluded e.g. from the temperature dependence of the Hall coefficient [\onlinecite{hwang1994}] and from the observation of a quadratic resistivity for overdoped \lsco\ [\onlinecite{nakamae2003}]. It is important to notice that two-particle scattering can only yield a finite resistivity, if momentum is transferred to the lattice, which only can be realized through umklapp scattering (see Sec. \ref{sec: Momentum relaxation mechanism in the cuprates and Matthiessens rule}). Umklapp scattering has special geometrical constraints and, therefore, introduces strong anisotropy in the Brillouin zone and requires a detailed analysis of the collision integral beyond a single-relaxation-time approach.

We linearize the Boltzmann equation in terms of the deviation from equilibrium $\delta f$. The collision integral is interpreted as a linear integral operator acting on $\delta f$. We evaluate the integral kernel on discrete patches of a finite mesh (see Fig.\ref{fig: discretization} (a)). Note in this context that that the energy conservation appearing as delta function in the scattering rates has to be treated carefully. In discretized momentum space, the linearized Boltzmann equation is reduced to a set of linear equations which are solved numerically.

In order to match the Fermi surface, we use polar angle and the energy to represent each momentum vector. For our curved-patch discretization we can adjust the grid spacing in radial (energy) direction to the temperature scale and always cover the relevant scattering phase space. A sketch of this discretization is given in Fig. \ref{fig: discretization} (a). For the calculations that we present here we have chosen a discretization of 24 patches in radial direction and 120 patches in angular direction. The latter are not distributed equally but are much finer in anti-nodal than in nodal direction to take care of the flat dispersion in the proximity of the saddle points.

\subsection{Calculation of the Seebeck coefficient within Boltzmann transport theory}
\label{subsec: calculation of the seebeck coefficient}

For a study of the Seebeck effect, the Boltzmann equation is extended by a new drift term that accounts for the response of the distribution function to the external thermal gradient,
\begin{align}
	-\frac{\partial f}{\partial \varepsilon({\bvec{p}})}\bvec{v}({\bvec{p}})\cdot\left(\bvec{\nabla}_{\bvec{r}}T\frac{\varepsilon({\bvec{p}})-\mu}{T} - \bvec{\mathcal E}\right)=\left.\frac{\partial f(\bvec{p})}{\partial t}\right|_\text{coll}
\label{eq: boltzmann equation with thermal gradient}
\end{align}
with the electro-chemical potential defined as $\bvec{\mathcal E}=\bvec{\nabla} (e\phi + \mu)$. The collision integral remains unchanged.

The distribution function that solves this equation carries a heat current and an electrical current. These two currents are defend as the linear response to the external fields as
\begin{align}
	\bvec{J}_{i}=\mathbf{K}_{ij}\bvec{F_{j}}, \quad \text{with}\quad
	\bvec{J}=\begin{pmatrix} \bvec{j}_{e} \\ \bvec{j}_{th} \end{pmatrix};
	\quad \bvec{F}=\begin{pmatrix}\bvec{\mathcal E} \\ -\bvec{\nabla}_{\bvec{r}}T\end{pmatrix}. 
\label{eq: J=KF - details}
\end{align}
Considering an open circuit geometry with $\bvec{j}_{e}=0$, the typical setup for thermopower measurement, the Seebeck coefficient is defined as the proportionality factor between the applied thermal gradient and the induced electrochemical potential, $Q\equiv\mathcal E / \nabla T=\text{K}_{12}/\text{K}_{11}$. For simplicity we assume all forces aligned along one direction and, thus, we can use scalar quantities. It is important to note that we do not make use of analytic simplifications as Mott's formula or Sommerfeld expansion.

\section{Conversion of computational to experimental units}
\label{sec: conversion of computation to experimental units}

According to the ARPES study [\onlinecite{yoshida2006}] used above to find the effective tight-binding parameters of \lsco, a value of $0.25\,\textrm{eV}$ for the nearest neighbor hopping parameter $t$ fits the ARPES data best. This defines an energy scale which translates the energy scales of temperatures in units of $t$ to $\textrm{eV}$. If we want to study the temperature range up to room temperature of $300\,\textrm{K}$ the corresponding energy scale is
$ k_{B}\,300\,\textrm{K} \approx 0.025\,\textrm{eV} \approx 0.1\,t $.
In order to cover at least one order of magnitude in temperatures we choose $k_{B}T_{\text{min}}=0.01\,t$ and $k_{B}T_{\text{max}}>0.1\,t$.

Our model is based on a two-dimensional interacting electron gas on a lattice. The experimental setup is based on the measurement of the resistivity of a three-dimensional but layered sample. For a quantitative comparison of the experimental and numerical data, a conversion of the two-dimensional resistivity to a three-dimensional resistivity is required. For a layered system this is very easily achieved by multiplication of the two-dimensional resistivity with the interlayer distance. The c-axis lattice constant of \lsco\ is given by $c=13.4\,\textrm{\AA}$ [\onlinecite{mourachkineBook}] and the interlayer distance by half the lattice constant, because the unit cell of the 2-1-4-compounds contains 2 layers.

\end{appendix}

\end{document}